\documentclass[twocolumn,times,twoside,10pt]{article}

\usepackage{times}

\usepackage[english]{babel}

\usepackage[a4paper,top=1.5cm,bottom=2cm,left=1.5cm,right=1.5cm,marginparwidth=0.75cm]{geometry}

\usepackage{amsmath}
\usepackage{amssymb}
\usepackage{latexsym}

\usepackage[round]{natbib}

\usepackage{graphicx}
\usepackage[colorlinks=true, allcolors=blue]{hyperref}

\usepackage{notation}

\title{A robust shape model for blood vessels analysis}

\newcommand{\corresp}{$^1$Corresponding Author $<$\href{mailto:ignacio.garcia@uv.es}{ignacio.garcia@uv.es}$>$. \\
Tel. +34 96 354 3064. Fax. +34 96 354 47 68.
}

\newcommand{\submittedto}{Article published in Applied Mathematics and Computation: 
\url{https://doi.org/10.1016/j.amc.2024.129078}}

\newcommand{\commlab}{CoMMLab -- Computational Multiscale Simulation Lab. University of Valencia. Spain.}

\author{Pau Romero$^a$, Abel Pedr\'os$^a$, Rafael Sebastian$^a$, Miguel Lozano$^a$ and Ignacio Garc\'\i{}a-Fern\'andez$^{a,1}$}
\date{}

\begin{document}

\twocolumn[
\maketitle

\begin{small}
\par\noindent $^a$\commlab
\end{small}

\begin{abstract}
The availability of digital twins for the cardiovascular system will enable insightful computational tools both for research and clinical practice. This, however, demands robust and well defined models and methods for the different steps involved in the process. 
We present a vessel coordinate system (VCS) that enables the unanbiguous definition of locations in a vessel section, by adapting the idea of cylindrical coordinates to the vessel geometry.
Using the VCS model, point correspondence can be defined among different samples of a cohort, allowing data transfer, quantitative comparison, shape coregistration or population analysis.
Furthermore, the VCS model allows for the generation of specific meshes (e.g. cylindrical grids, ogrids) necessary for an accurate reconstruction of the geometries used in fluid simulations.
We provide the technical details for coordinates computation and discuss the assumptions taken to guarantee that they are well defined.
The VCS model is tested in a series of applications. We present a robust, low dimensional, patient specific vascular model and use it to study phenotype variability analysis of the thoracic aorta within a cohort of patients. Point correspondence is exploited to build an haemodynamics atlas of the aorta for the same cohort. 
The atlas originates from fluid simulations (Navier-Stokes with Finite Volume Method) conducted using OpenFOAMv10.
We finally present a relevant discussion on the VCS model, which covers its impact in important areas such as shape modeling and computer fluids dynamics (CFD).
\end{abstract}


\par\rule{0.45\textwidth}{1pt}

\begin{small}
\noindent
\corresp\\
\submittedto
\end{small}

\vspace{1cm}
]



        \section{Introduction}
        \label{sec:intro}

Encoding and representation of anatomy is
one of the grounding pieces for the development of the so-called digital twin~\citep{corral-acero20}. 
When building a digital patient, data from different sources has to be put together 
on a common anatomical substrate~\citep{bayer12,gil19,lopez-perez19, salmasi21}.
The combination of mechanistic models with data science models has proven to be a powerful tandem
for interpreting simulation results, generating data for model training, or synthesizing large virtual cohorts for in-silico trials~\citep{alber19,baker18,geronzi23,lamata18,niederer20,romero21a,thamsen21}.
These computational tools require suitable representations of the anatomy to encode patient information, which means representations that are independent of the patient from a computational point of view~\citep{schuler21}.

In the case of modeling vascular anatomy, we find different technical solutions and strategies that we can classify in four groups.
Some authors use a set of anatomical traits that are relevant for their clinical problem.
In order to study characteristics associated to ascending aorta aneurysm,~\cite{cosentino20} define a reduced set of 15 anatomical features and relate them with biomechanical descriptors biomarkers. This approach is not intended, however, to provide an encoding of the anatomy beyond the goal of the study, and it does not allow, for instance, the reconstruction of an aorta from its characteristics.
A common approach for a computational description of vessels is to extract their centerline and analyze the morphology of cross sections. Following this approach,~\cite{antiga04,antiga08} build models for bifurcating vessels by treating their lumen as the union of spheres located along the centerline. \cite{medrano-gracia16}~build an atlas of vessel bifurcations also by studying the centerline of a vessel tree and their cross sections at regular distances. In order to have an analytic description of the wall of a vessel, some authors sweep closed curves, such as ellipses~\citep{alvarez17} or polynomials~\citep{urick19,zhang07}, along the centerline. 
A more general approach, in the sense that it is not specific for vessels, is the use of \textit{large deformation diffeomorphic metric mappings}~\citep{durrleman09} to encode the geometry. This technique characterizes geometries through the integral of vector fields on their surface, and has been successfully applied to study the variability of aortic arc shape~\citep{bruse17} but also to analyze other organs such as the whole heart~\citep{rodero21}.
When machine learning models are to be involved, it is often a requirement that all the samples use a common encoding; the same dimensionality, with common meaning for all features. With this aim, Liang and coauthors use a reference triangle mesh for the wall of the thoracic aorta and project its points onto the wall of each sample. By doing so, all the aortas are described by a template, adapted to the patient specific phenotype. The template is, then, fed into a neural network to estimate functional biomarkers of the aorta~\citep{liang17,liang18a,liang20}.

This diversity of methods enable descriptions tailored to the requirements of each problem. However, not all methods are equally adequate for inter-patient comparison or for population based statistical studies. Furthermore, this diversity of approaches might hamper the reproducibility and the comparability of different research works.
For these reasons, some authors are stressing the importance of having robust and standardized methods for all the phases of the development and analysis of multiscale biophysical models and simulation results~\citep{bayer18,niederer20,roney19,schuler21,urick19}.

In this paper, we propose a new representation of the anatomy of blood 
vessels. We present a Vessel Coordinate System (VCS) that serves as a 
reference to describe a location on the vessel wall or in its lumen.
The VCS enables the quantitative comparison 
of data fields defined on a set of vessels corresponding to the same 
anatomical section, and serves as a tool for the analysis and 
visualization of fields defined on a vessel.
We provide a detailed description of the different steps to build the reference system, and justify the particular decisions, with the aim of reproducibility, anatomy 
comparability, and robustness. We show that, on the proposed VCS, a robust, low dimensional, differentiable representation of the vessel surface can be built, enabling point-wise 
comparison of anatomies and of data fields defined on them, without
ambiguity and at a very reasonable computational cost. After defining the VCS in Section~\ref{sec:mat_meth}, we show some use cases. In Section~\ref{ssec:PSVM_residuals} a patient specific vascular model for the thoracic aorta is built from medical image data. In Section~\ref{ssec:SSM_results} a statistical shape model of the aorta is developed from a cohort of patients, and, finally, in Section~\ref{ssec:haemo_atlas_res} an haemodynamic atlas is computed for that cohort.



        \section{Material and methods}
        \label{sec:mat_meth}

    We have developed a coordinate system that enables a unique description of a point inside the vessel lumen or on the vessel wall. 
    Our scheme is similar to the one presented in~\citep{bayer18} to define a coordinate system on the ventricle.

    \subsection{Material}
    \label{ssec:mat}

\begin{figure*}[t!]
        \centering
        \includegraphics[width=.95\textwidth]{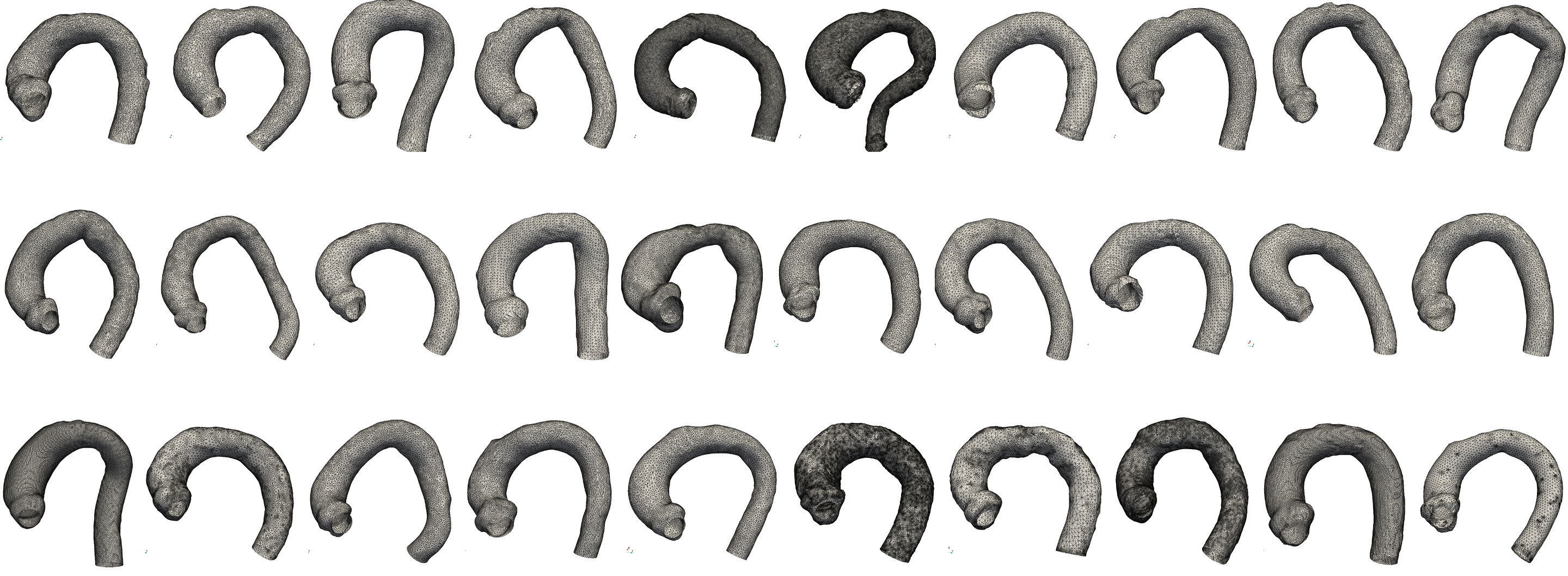}
        \caption{Saggital view of the 30 thoracic aorta meshes used for the study.}
        \label{fig:mat_meshes}
    \end{figure*}

    We have used a data set of 30 thoracic aortas, to test our methodology on a wide range of anatomies with variable ratios of curvature and diameter.
    The set of anatomies corresponds to a cohort of 30 patients with ages ranging from 78 to 89 years old, suffering from aortic valve stenosis.
    {The cohort has variability of ascending aorta diameters, with values ranging from around 3,5cm to over 6cm in some patients with aneurysm.}
    Expert radiologists manually segmented the aortas from computerized tomography images, acquired at the mesosystole phase of the cardiac cycle. 
    Since the VCS is restricted to single branch vessels, the supra-aortic branches as well as the coronary arteries were removed in all the cases. 
    The final input data set used in this study was the set of 30 anonymized triangular surface meshes (see Fig.~\ref{fig:mat_meshes}). All the data acquisition process met the requirements of the Declaration of Helsinki and was approved by the institutional ethics committee. 
    %

    \subsection{Vessel coordinate system}
    \label{ssec:VCS_def_and_comp}

    The wall of a vessel section --with no bifurcations-- can be considered as homeomorphic (i.e. can be transformed through a continuous bijection) to a cylinder. 
    The proposed VCS is a generalization of the cylindrical coordinate system applied to the vessel,
    as in~\citep{bersi16, bersi19, meister20, rego21}, where they align the longitudinal axis of the vessel with the \textit{z} axis (corresponding to  \textit{height}) of the cylindrical reference system. However, for this alignment to be possible, a longitudinal axis must be defined, which limits the application of cylindrical coordinates to almost straight vessel sections.

    A natural generalization to this scheme is to substitute the longitudinal axis by a regular parametric curve inside the vessel, and to define a local reference frame at every point of the curve. Then, a location $\xx\in\bbbr^3$ inside the vessel's lumen can be identified by the longitudinal location of the closest point of the curve to $\xx$ and by the coordinates of $\xx$ in the local reference frame corresponding to that closest point.
    We describe how this coordinate system can be built so that both the curve and the local reference frames are properly defined in a wide range of situations.

    
    \subsubsection{Definition of the vessel coordinate system}
    \label{sssec:outline_vcs}
    
\begin{figure}[!ht]
    \centering
    \includegraphics[width=\linewidth]{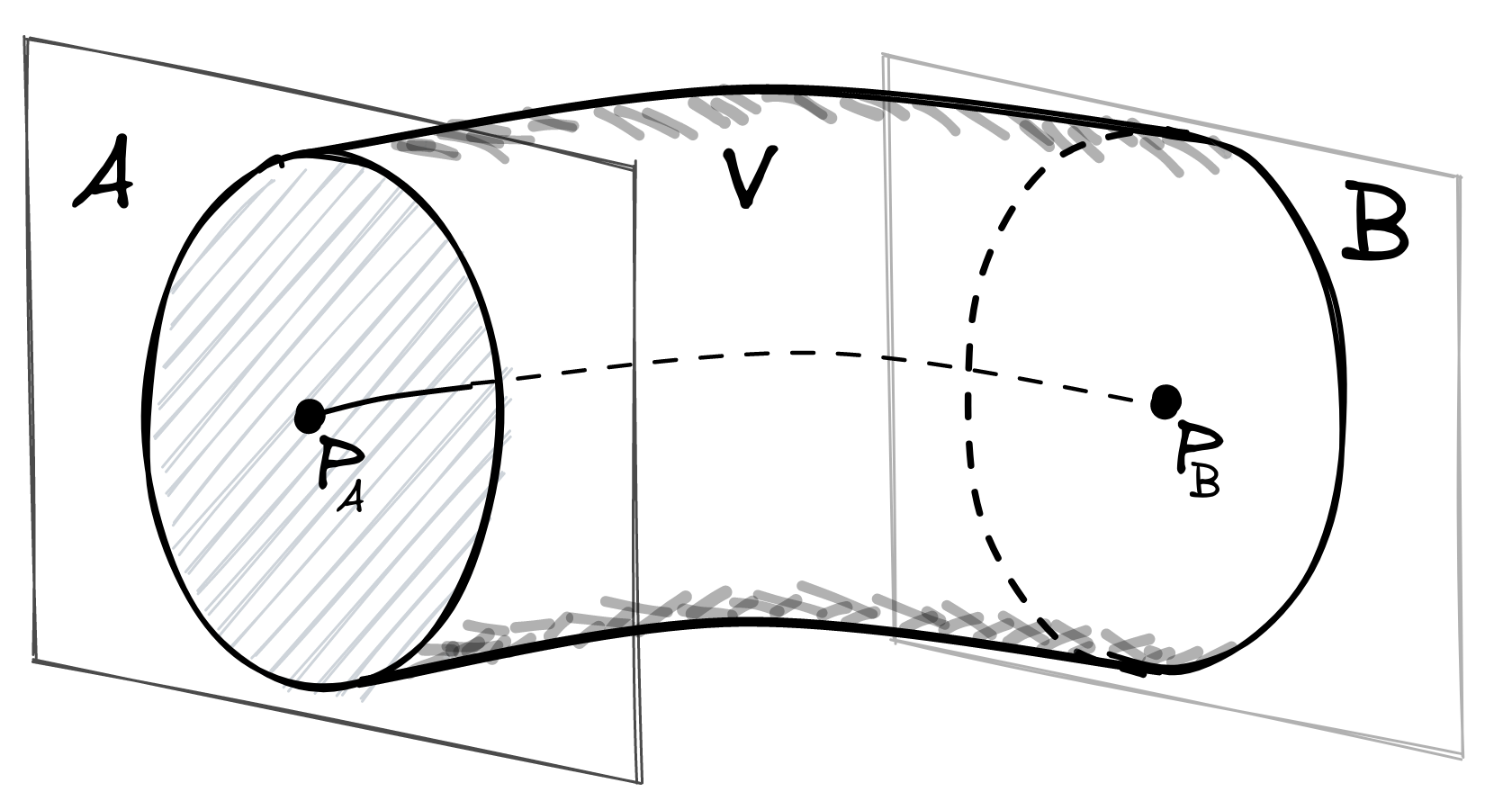}
    \includegraphics[width=0.85\linewidth]{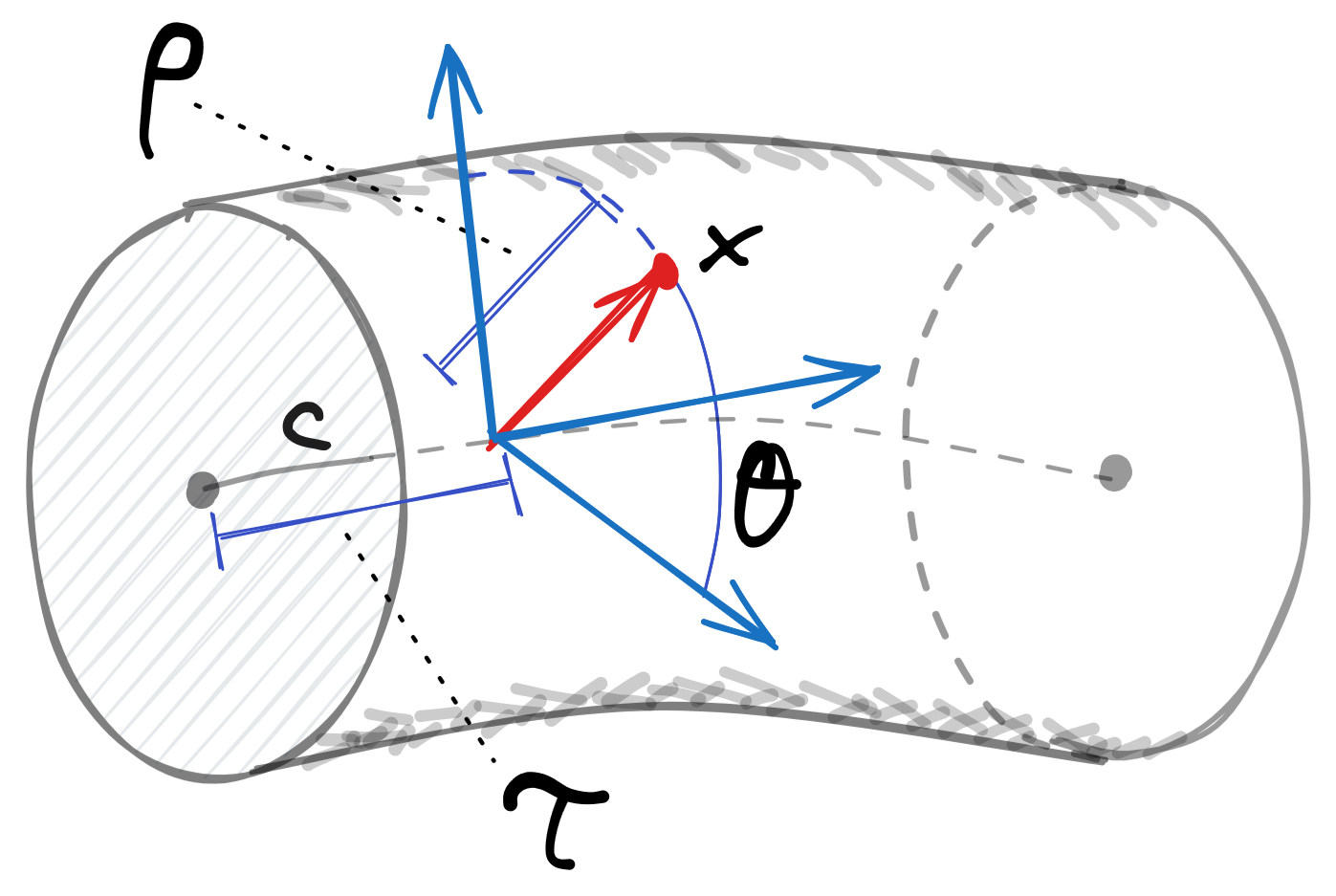}
    \caption{The Vessel Coordinate System is defined on a vessel section $V$, confined by two cross sections $A$, and $B$ (top). The longitudinal axis of the cylinder is represented by a curve, $\cc$, on which a local reference frame is defined, $\left\{\bt(t),\vv_1(t),\vv_2(t)\right\}$ (bottom figure, arrows in blue), with $\bt$ the unitary tangent to~$\cc$. Given a point $\xx$, its coordinates are its longitudinal location along the curve, $\tau$, and the polar coordinates of~$\xx$ in the plane orthogonal to the curve containing $\xx$, $(\rho,\theta)$, represented in red in the figure.}
    \label{fig:vcs_def}
\end{figure}

    We define the VCS for a vessel segment that is limited by two cross sections. We assume that the region of interest has already been segmented, and that we have identified the location of the wall, either as a triangle mesh or as a volumetric representation. We assume that the two ends of the vessel section are defined by two planes $A$ and $B$ approximately orthogonal to the vessel direction and that we have two points $\pp_A\in A$ and $\pp_B \in B$ inside the vessel lumen, as shown in Fig.~\ref{fig:vcs_def}. 

    For this segment, we start by building a regular parametric 
    curve, $\cc:I\to\bbbr^3$, that defines the longitudinal direction of the vessel segment, and a local reference frame defined on the curve, $\left\{\bt(t),\vv_1(t),\vv_2(t)\right\}$, with $\bt$ the unitary tangent to $\cc$. Then, given a point $\xx\in\bbbr^3$ on the wall or inside the lumen of the vessel, we define its cylindrical coordinates as
    \begin{align}
        \tau &= \argmin_{t\in I} \{\|\xx-\cc(t)\|\};
        \label{eq:vcs_tau} \\
        \theta &= \ang_{\bt}\left(\vv_1(\tau),\xx - \cc(\tau) \right),
        \label{eq:vcs_theta} \\
        \rho &= \|\xx-\cc(\tau)\|;
        \label{eq:vcs_rho}
    \end{align}
    where the function $\ang_{\bt}(\cdot,\cdot)$ gives the angle between two vectors, rotating around the direction defined by $\bt$.
    The bottom diagram in Fig.~\ref{fig:vcs_def} shows the coordinate system for a point inside the vessel.
    The parameter $\tau\in I$ represents the longitudinal position of $\xx$ in the vessel segment, 
    while $\theta\in[0,2\pi]$ and $\rho\in\bbbr^+$ are the polar coordinates of $\xx$ in the reference frame $\left\{\vv_1,\vv_2\right\}$ defined at $\cc(\tau)$.
    Given three values $\tau\in I$, $\theta\in[0,2\pi]$ and $\rho\in\bbbr^+$, they determine a point in $\bbbr^3$ that meets the previous definition by means of
    \begin{equation}
    \label{eq:param_tube}
    \mathbf{x}(\tau, \theta, \rho) = \cc(\tau) + \rho (\vv_1 \cos(\theta) + \vv_2 \sin(\theta)).
    \end{equation}

    For this coordinate system to be well defined, the longitudinal curve, $\cc$, and the local reference frame, $\left\{\vv_1(t),\vv_2(t)\right\}$, have to be, in turn, properly defined. In this section we describe the basic properties each piece of the VCS needs to have, and we explain how to compute them meeting these requirements.


        
        \subsubsection{Computation of the longitudinal curve}
        \label{sssec:centerline_computation}

        The first step to compute the VCS is to compute the curve that determines the longitudinal direction of the vessel segment. We compute this curve as the centerline or medial axis of the geometry~\citep{blum67}.
        In our work we build the centerline as the curve $\cc$ that connects the two planes $A$ and $B$ and that maximizes the minimum distance from the vessel wall to every point of $\cc$.

        First, on a volumetric discretization of the vessel lumen, which can be obtained directly from medical image data or generated from the wall points, we compute the distance field with respect to the wall. Then, we build a discrete path that traverses the lumen points maximizing the distance to the wall along the curve. In order to do this, we use the A$^*$ graph search algorithm, taking the inverse of the distance scalar field as the cost function, and a neighborhood defined by distance~\citep{romero22}.
        We finally build a differentiable curve by adjusting cubic B-spline to the  set of points that form the discrete path. Without loss of generality, we will take a parametrization of the curve with constant velocity and $I=[0,1]$.


        \subsubsection{Local reference frame}
        \label{sssec:pa_frame}
        The definition of the angular coordinate, $\theta$, requires a stable angular origin in the normal plane along the centerline. Moreover, if comparison between different patients' vessels is one of our goals, we need to provide stability also among the different anatomies of a cohort. A reference frame based on the intrinsic geometry of the curve, such as the Frenet frame, would not provide such stability and could lead to undetermined directions (e.g. where the curvature vanishes)~\citep{piccinelli09}.
        Instead, we rely on the definition of an arbitrary reference frame at the beginning of the curve $\cc(0)$, together with its parallel transport along the curve. 
        The parallel transport of a vector orthogonal to a curve sweeps the vector along the curve in a way that it changes the minimum required to keep it inside the normal plane~\citep{guo13}.
        More formally, given a parametric curve $\cc:[0,1]\to\mathbb{R}^3$, and a vector $\vv^0 \perp \bt(0)$, the vector field $\vv(s)$ is the parallel transport of $\vv^0$ if its derivative w.r.t. $s$ can be entirely expressed in terms of the tangent vector, $\bt$, i.e. $\dot\vv(s) = g(s)\bt(s)$ for some function $g(s)$.
        The interested reader will find further detail in~\citep{bishop75, carroll13}.

        In order to build a stable reference frame along the centerline, we start by defining a reference frame at the initial point of the curve, $\left\{\vv_1^0,\vv_2^0 \right\}$, and compute its parallel transport, $\left\{\vv_1(s),\vv_2(s) \right\}$, using a forward Euler integration scheme as follows. Assuming that the parallel transport has been computed up to $s\ge0$, and given $h>0$, we compute the frame at $s+h$ by rotating $\vv_1(s)$ and $\vv_2(s)$ by an angle $\alpha = \arccos(\bt(s)\cdot \bt(s+h))$
        around the direction defined by $\rr = \bt(s)\times \bt(s+h)$.
        From now on, we will refer to the local reference frame at a point of the curve $\cc(s)$ as $\left\{\vv_1,\vv_2 \right\}$, dropping the parameter $s$ unless it leads to ambiguity.

        \subsubsection{Computation of the coordinates for a point}
        \label{sssec:comput_coords}

        Next, we describe the procedure we use to compute the coordinates of a given point $\xx\in\bbbr^3$. For now, we assume that the point $\xx$ is such that the minimization problem defined by Eq.~(\ref{eq:vcs_tau}) has a unique solution. Later on, in Section~\ref{sec:discussion}, we will discuss why this is a reasonable assumption and how to overcome situations in which this assumption could fail to be true.

        The first step is to define the reference frame for the initial point of the curve, $\left\{\vv_1^0,\vv_2^0 \right\}$. The election of this reference frame is arbitrary and should be done taking in consideration the particular goals of the study, e.g., in all our tests, we take $\vv_1^0$ so that the centroid (the average) of the points of the wall is contained in the plane defined by $\bt(0)$ and, $\vv_1^0$.

        \paragraph{Longitudinal coordinate, $\tau$.}
        In order to compute $\tau$ we need to solve Eq.~(\ref{eq:vcs_tau}), which stands for finding the closest point $\cc(\tau)$ to $\xx$ on the curve $\cc$.
        The optimality condition for that problem states that the tangent to the curve, $\bt$, and the vector $\cc(\tau) - \xx$ must be orthogonal. Thus, we can directly minimize Eq.~(\ref{eq:vcs_tau}) or solve the nonlinear equation $\bt(\tau)\cdot(\cc(\tau) - \xx) = 0$.
        In either case, we can benefit from the fact that $\cc$ is an order 3 polynomial for which we have its derivatives analytically.

        \paragraph{Local polar coordinates, $\rho$, $\theta$.}
        Once $\tau$ has been computed for $\xx$, the value of the radial coordinate, $\rho$, which stands for the distance from $\xx$ to the curve $\cc$, is immediately obtained as $\rho = \|\cc(\tau) - \xx\|$.
        The computation of the angular coordinate, $\theta$, requires the local reference frame $\left\{\vv_1(\tau),\vv_2(\tau) \right\}$. This reference frame is obtained, as described previously, by solving the parallel transport of the initial reference frame $\left\{\vv_1^0,\vv_2^0 \right\}$ from $\cc(0)$ to $\cc(\tau)$. The value of $\theta$ is computed from the dot product of $\cc(\tau) - \xx$ with the vectors $\vv_1$ and $\vv_2$ as the angle from $\vv_1$ in the range $[0,2\pi]$. Note that, by construction, all three vectors are in the plane orthogonal to $\bt(\tau)$ that contains $\cc(\tau)$.

        \subsection{Encoding vessel anatomy}
        \label{ssec:geom_enc}

        After defining the VCS, we will present several contexts in which it can be applied. In the first place, we discuss how the coordinates defined on a vessel section can be used to build a representation of the vessel itself. We address the development of patient specific models that are based on a common representation of the wall, making it easy to establish a correspondence between two individuals. We also benefit from the existence of this correspondence to build a convenient representation of a cohort of patients.

        \subsubsection{Patient specific vascular model}
        \label{sssec:PSVM}

        We start the process with segmented medical image data of the vessel and, more precisely, with a set of points on the vessel wall.
        We will take the common assumption that the vessel wall can be represented as a differentiable surface in the region of interest. Furthermore, we also assume that any section of the vessel, perpendicular to the vessel centerline, is a 
        {star-convex} set: that is, all the points in the cross section can be seen from the centerline. 
        Although this assumption imposes a limitation in some anatomical conditions, such as saccular aneurysms, it is still valid in a wide range of situations.
        Under this assumption, the radius of any point on the wall $\rho(\xx)$ can be expressed as a function of the other two coordinates of that point, $\rho(\xx) = \rho_w(\tau(\xx),\theta(\xx))$. 
        Then, using Eq.~(\ref{eq:param_tube}), the wall can be described as the surface defined by
        \begin{equation}
        \label{eq:param_wall}
        \mathbf{x}(\tau, \theta) = \cc(\tau) + \rho_w(\tau,\theta) (\vv_1 \cos(\theta) + \vv_2 \sin(\theta)),
        \end{equation}
        for $\tau\in[0,1],~\theta\in[0,2\pi]$. Since the reference frame $\{\vv_1,\vv_2\}$ is defined for any $\tau$ from $\cc$, in order to represent the anatomy of a patient it is enough to have a representation of $\cc$ and of $\rho_w$.
        Our proposal is to use a differentiable approximation by fitting a cubic spline curve for $\cc$ and a bivariate spline function for $\rho_w$.

        We approximate the centerline $\cc$ as
        \begin{equation}
            \cc(\tau) = \sum_{i=0}^n\cc_iB_{\bt,i}(\tau)
            \label{eq:c_bsp_fit}
        \end{equation}
        where $\cc_i$ are 3-dimensional coefficients and $B_{\bt,i}$ are a basis of the space of B-Spline polynomials.
        The knot vector, $\bt\in [0,1]^{n+5}$, is a partition of the interval $[0,1]$ and defines the region of influence of each $B_{\bt,i}$. We take $\bt$ as a uniform subdivision of $I=[0,1]$, with $t_k = k/(n+4),~k = 0,\ldots,n+4$, leading to so-called \textit{uniform B-Splines}. From now on, vector $\bt$ will be assumed uniform and omitted in the notation for simplicity. The interested reader is addressed to the work by~\cite{piegl96} for a detailed description of B-Splines.
        The approximation to the centerline is computed as the solution to a least squares problem on the coefficients $\cc_i$, to minimize the difference from $\cc$ to the centerline points defined in Section~\ref{sssec:centerline_computation}.
        
        For the wall, we get a sample of points 
        on the surface of the vessel, and fit a uniform bivariate spline
        \begin{equation}
            \rho_w(\tau,\theta) = \sum_{i=0}^{n_1}\sum_{j=0}^{n_2}b_{ij}B_{ij}(\tau,\theta)
            \label{eq:rho_bsp_fit}
        \end{equation}
        where $B_{i,j}(x,y) = B_{i}(x)\cdot B_{j}(y)$ and $(\tau,\theta)\in [0,1]\times[0,2\pi]$.
        
\begin{figure*}[t!]
            \centering
            \includegraphics[width=0.95\linewidth]{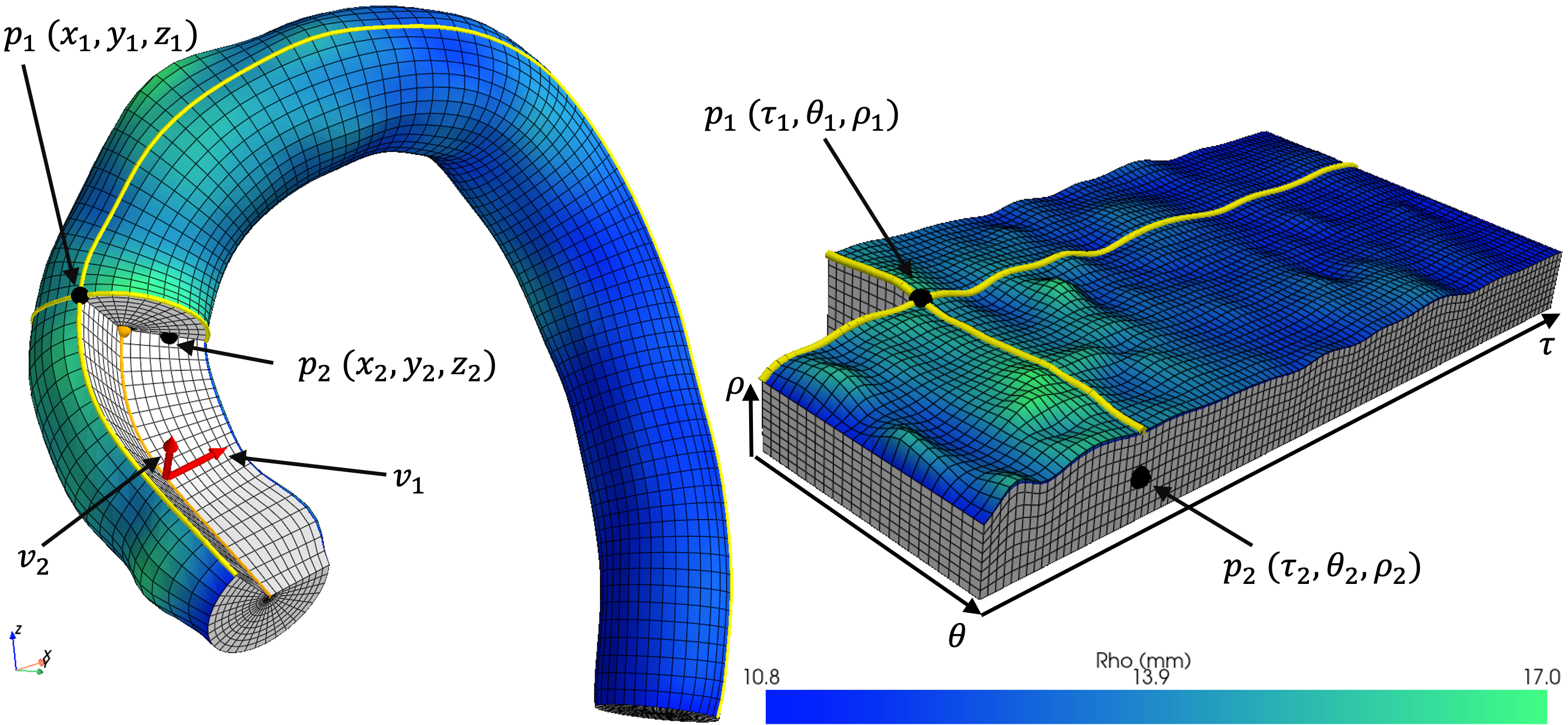}
            \caption{
            The representation of an aorta using the VCS. After computing the centerline of the vessel (in orange), the wall is approximated as a surface $\rho_w$ depending on coordinates $(\tau,\theta)$. This surface is represented, on the right, in VCS space. The color map corresponds to the distance of the wall points to the centerline, in millimeters, 
            {with the radius ranging, for this aorta, from $\sim$10mm to $\sim$17mm}.
            Two points, $p_0$ (on the wall) and $p_1$ (inside the lumen) are represented in both Cartesian and VCS spaces, as well as two lines on the surface (in yellow), corresponding to $\theta=5\pi/4$ and $\tau=1/5$. The reference frame on the centerline is also shown in red in the Cartesian space for a point of the centerline.
            }
            \label{fig:rho_func}
        \end{figure*}

        Finally, by using these spline approximations in Eq.~(\ref{eq:param_wall}), we represent the aorta wall as
        %
        \begin{small}            
        \begin{equation}
        \label{eq:param_wall_bsplines}
        \mathbf{x}(\tau, \theta) = \sum_{i=0}^n\cc_iB_i(\tau) + 
        \left[\sum_{i=0}^{n_1}\sum_{j=0}^{n_2}b_{ij}B_{ij}(\tau,\theta)\right]
        (\vv_1 \cos(\theta) + \vv_2 \sin(\theta)).
        \end{equation}
        \end{small}
        Fig.~\ref{fig:rho_func} shows the application of this approach to an aorta in our dataset.

        This approach leads to a differentiable representation of the anatomy that uses a reduced number of real values; given values for $n,n_1,n_2$, 
        any vessel section is encoded using a vector with fixed dimension $m = 3n+n_1\cdot n_2$, with the form
        \begin{equation}
            \ba = (\cc_0,\ldots,\cc_n,b_{00},\ldots,b_{n_1n_2}) = (a_1,\ldots,a_{m}).
            \label{eq:feature_vector}
        \end{equation}
        
        Once the vessel has been characterized through the corresponding feature vector $\ba$, an approximation of its original shape can be recovered by means of Eq.~(\ref{eq:param_wall_bsplines}). The quality of this approximation will depend on the number of knots used for the different splines. 
        To assess the quality of this approximation of the original wall, we will compute the patient specific model for all the aortas in our test cohort using different knot vectors, and compare the approximated wall with the original data.
        Given a point of the wall with Cartesian coordinates $\pp = (x,y,z)$, we will compute its vessel coordinates $(\tau(\pp),\theta(\pp),\rho(\pp))$, and define the residual of the approximation as
        \begin{equation}
        \label{eq:residuals}
            r(\pp) = \left|\pp  - \mathbf{x}(\tau(\pp),\theta(\pp))\right|.
        \end{equation}

        \subsubsection{Coregistration of a cohort of vessels}
        \label{ssssec:GPA}
 
        A common preprocessing step when dealing with several samples of the same anatomy is geometry alignment or coregistration. 
        Coregistration methods often rely on the alignment of a set of points defined on the different anatomies. 
        Ideally, there should be a one to one correspondence among these points on the different samples. However, in many cases it is not easy to define this correspondence beforehand, and specific numerical methods, such as Iterative Closest Point, are used.
        The patient specific model proposed here has two properties that make it useful for geometry coregistration. 
        
        In the first place, the construction of the VCS for every patient enables the definition of a set of equivalent alignment points on all the geometries. E.g., one can take points on the surface of the vessel forming a regular grid in the $\tau-\theta$ domain, with an automatic correspondence by definition.
        In the second place, the coefficients of the centerline have a geometrical meaning as control points. As a result, the application of a rigid transformation on these points results on the application of the transformation to the whole vessel anatomy.

        Thus, when dealing with a set of geometries that require rigid coregistration, this step can be assisted by building the vascular models of all the anatomies before the alignment. The alignment can then be performed on a set of points built for this task with an explicit one to one correspondence, thanks to the vessel coordinate system. During the process, the resulting transform can be applied only to the centerline control points, without requiring the recomputation of the vessel model after every step.

        \subsubsection{Population representation and analysis}
        \label{ssssec:SSM}
        {The set of functions that can be expressed by Eq.~(\ref{eq:param_wall_bsplines})} form a vector space, isomporhic to the space of coefficient vectors defined in Eq.~(\ref{eq:feature_vector}). This 
        can be exploited to represent a particular phenotype as the combination of a reference shape plus a perturbation or deformation, making it very convenient for methods such as statistical shape modeling.
        Here, this idea will be developed by generating a population model on our cohort, using Principal Component Analysis (PCA).

        If we have $M$ vessel geometries, represented by their coefficients
        \begin{small}
        $$
        \ba^k = (\cc_0^k,\ldots,\cc_n^k,b^k_{00},\ldots,b_{n_1n_2}^k) = (a_1^k,\ldots,a_{m}^k),~~k=1,\ldots,M,
        $$
        \end{small}
        as defined in equations (\ref{eq:param_wall_bsplines}) and (\ref{eq:feature_vector}), 
        each geometry can be represented by the mean vessel geometry, $\bmu=\sum_j\ba^j/M$, plus the projection of its difference with $\bmu$ on the un correlated principal directions, $\uu_i$
        \begin{equation}
        \label{eq:PCA}
            \mathbf{a} = {\bmu} + \sum_{i=0}^{m'} \alpha_i \uu_i.
        \end{equation}
        In this context, in which each $\uu_i$ can be interpreted as a change in shape that increases with $\alpha_i$, the principal directions are also called variation or deformation modes.

        We will compute the PCA of the aortas in our cohort and will explore the meaning of each deformation mode, to understand how the variation in shape within the cohort can be explained. By sampling the aortas' geometry at a set of fixed points, defined by their coordinates, we will quantify the deformation both of the centerline and of the width of the aorta under each deformation mode.

    \subsection{Application to inter-patient variability analysis}
    \label{ssec:field_comp}

        The VCS defines a one to one correspondence between points on two vessels, enabling the quantitative comparison of any field that is defined on its surface or inside its lumen.
        We will show this approach building an haemodynamic atlas based on Computational Fluid Dynamics (CFD) simulations on the cohort of aortas. For a quantitative comparison, we will define a common set of measurement points $\xx_i$ through their coordinates, and estimate on them the value of different quantities for all the vessel samples.

        \subsubsection{CFD simulation setup}
        \label{sssec:CFD_setup}

        For our computational blood flow simulations we will follow the approach by~\cite{catalano21}, 
        {considering rigid walls and steady state simulations with constant inflow boundary conditions in peak systole}. We will assume the blood to be an incompressible, Newtonian fluid in laminar regime, with a kinematic viscosity of $3.37 \cdot 10^{-6}$  m$^2$/s. In all the aortas, we will impose comparable boundary conditions; a constant Dirichlet condition for the inlet velocity profile equal to 1.2 m/s, with an effective valve area of 1 cm$^2$ corresponding to a patient with aortic stenosis according to~\cite{vahanian21}; the standard no-slip condition, i.e. zero velocity Dirichlet condition, for the wall; and free outflow with null von Neumann boundary condition for the velocity at the outlet on the descending aorta.
        The open source framework OpenFOAM~\citep{openfoamfoundation}, version 10, will be used to solve the Navier-Stokes equations by means of the Finite Volume Method (FVM). We will generate the aorta wall in the form of a regular triangle mesh, using Eq.~(\ref{eq:param_wall_bsplines}). This triangle mesh will be used as the input of the snappyHexMesh software provided with OpenFOAM, to generate hexa-dominant volume computational meshes. The SIMPLE algorithm will be used to approximate the solution, assuming convergence when residuals fall below $10^{-5}$.
        It is noteworthy that the simulations have not been physiologically validated, since their purpose is to serve as a use case of the VCS. Hence, no biophysical conclusions should be taken from the results.

        \subsubsection{Haemodynamic atlas on the aorta}
        \label{sssec:haemo_atlas}
        {We will build the haemodynamic atlas following an approach similar to the analysis of anatomy in Section~\ref{ssssec:SSM}. 
        We perform a CFD simulation for each of the vessels in the cohort and, in this case, we compute a PCA for each one of the flow variables across the cohort.}
        This strategy is usually hindered by the fact that the computational grids of two vessels cannot be compared directly. Moreover, it is not straightforward to define measurement points on which perform the desired comparison.
        By means of the VCS, a comparable measurement grid will be generated for all the aortas, considering the correspondence defined by the coordinate values.
        To ease the notation, we will normalize the radial coordinate; given a point $\xx$ with coordinates $(\tau,\theta,\rho)$, we define 
        $\rho_n (\xx)= \rho / \rho_w(\tau, \theta)$.
        This normalized coordinate is always in the range $[0,1]$ for a point inside the aorta, with $\rho_n=0$ still representing points located on the centerline, while $\rho_n=1$ now corresponds to points on the wall. Then, we will build a regular grid in the region $\Omega = [0,1]\times[0,2\pi]\times[0,1]$ and use the resulting points to measure the quantities of interest on the resulting points of every aorta from the computed CFD solution.
        As a result, every aorta will have a comparable vector of measurements that allows quantitative comparison.

        For our atlas, we will focus on pressure, velocity and wall shear stress (the later only defined for points on the wall). 
        For each one of these fields, a separate PCA has been performed; the average dynamics will be computed within the cohort, and the principal uncorrelated modes of variation of the flow will be studied, showing the most significant differences observed.


\section{Results}
\label{sec:res}
    Next we present the results of the different experiments proposed. Following the structure of Section~\ref{sec:mat_meth}, we start with the analysis of the patient specific model as a tool to approximate the vessel wall. Then, we present the results of the statistical shape analysis of the available cohort, and conclude with the construction of an haemodynamic atlas for this set of aortas.

    \subsection{Patient specific model}
    \label{ssec:PSVM_residuals}

        As described in Section~\ref{sssec:PSVM}, the parametric model of a vessel defined by Eq.~(\ref{eq:param_wall_bsplines}) depends on 
        the size of the knot vectors, that control the degrees of freedom of the B-Splines. 
        We will denote by $L$ the number of knots for $\tau$ in the approximation of $\cc$. $K$ and $R$ will be, respectively, the number of knots for $\tau$ and $\theta$ in the approximation of $\rho_w$. 
        The quality of the approximation of the parametric surface given by Eq.~(\ref{eq:param_wall_bsplines}) has been measured for each aorta using the residual $r$ defined in Eq.~(\ref{eq:residuals}).
        This results in a distribution of $r$ for each aorta across its surface.

\begin{figure*}[ht!]
\centering
\includegraphics[width=.3\linewidth, trim={0cm 0cm 0cm 0cm}, clip]{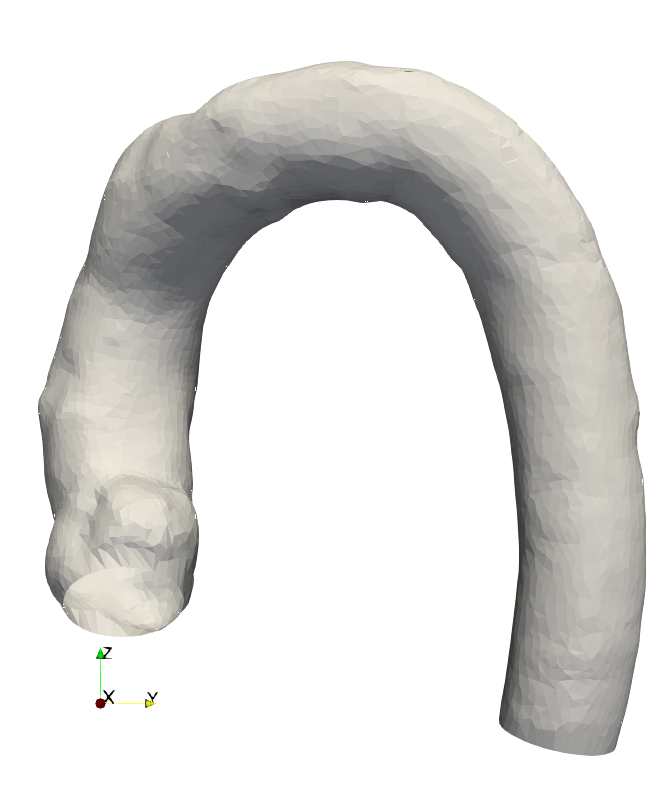}
\hfill
\includegraphics[width=.3\linewidth, trim={0cm 0cm 0cm 0cm}, clip]{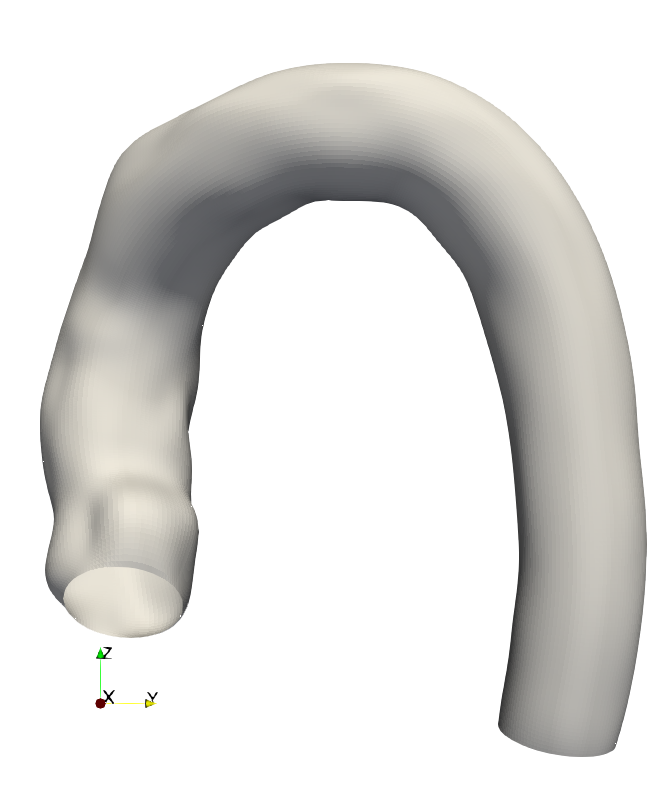}
\hfill
\includegraphics[width=.3\linewidth, trim={0cm 0cm 0cm 0cm}, clip]{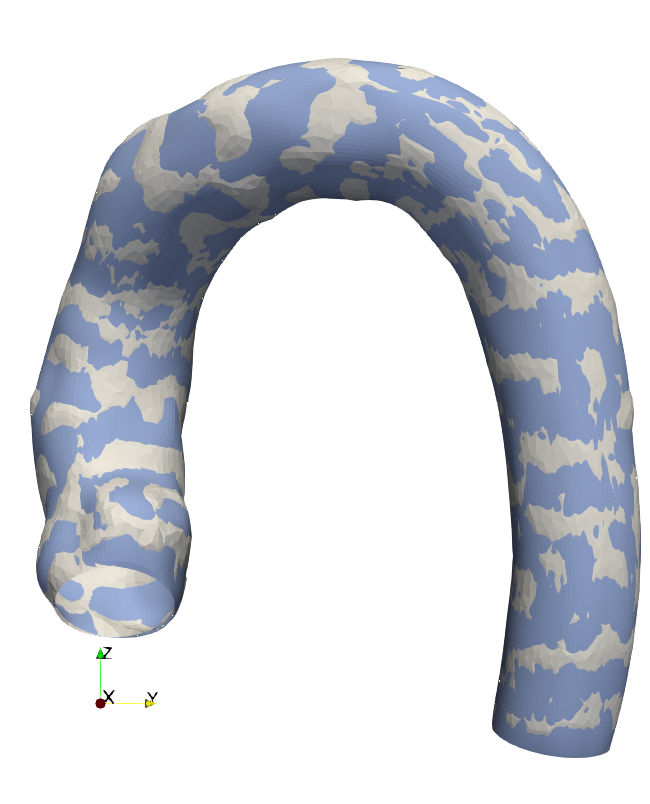}\\
\includegraphics[width=.3\linewidth, trim={0cm 0cm 0cm 0cm}, clip]{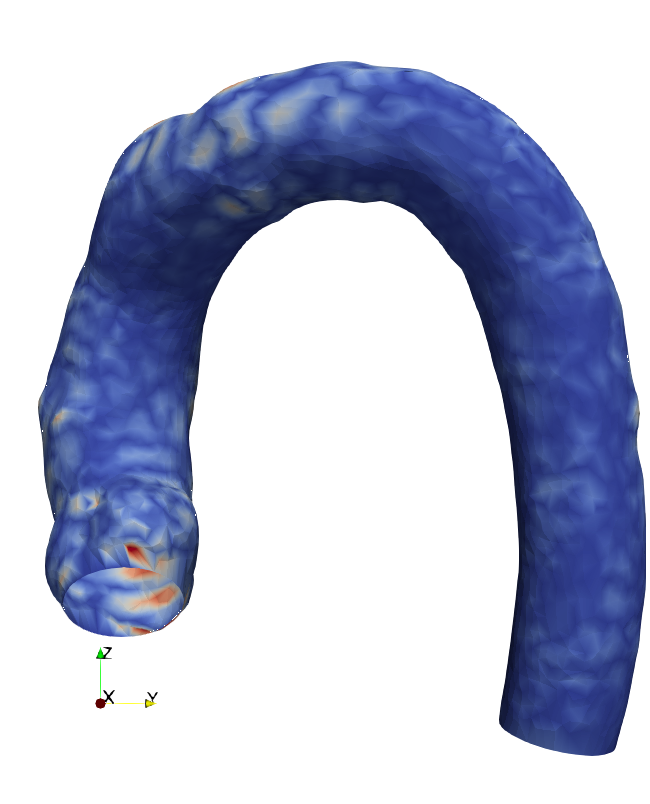}\hfill
\includegraphics[width=.3\linewidth, trim={0cm 0cm 0cm 0cm}, clip]{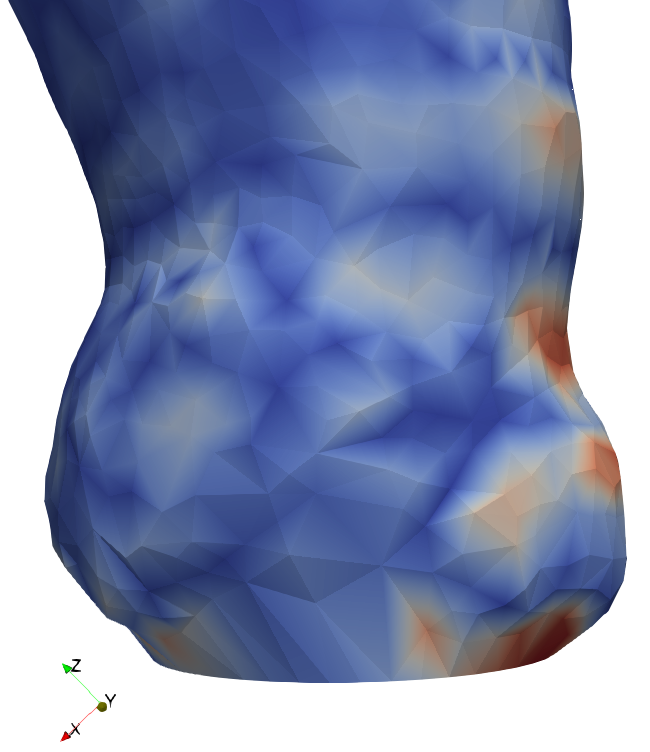}\hfill
\includegraphics[width=.3\linewidth, trim={0cm 0cm 0cm 0cm}, clip]{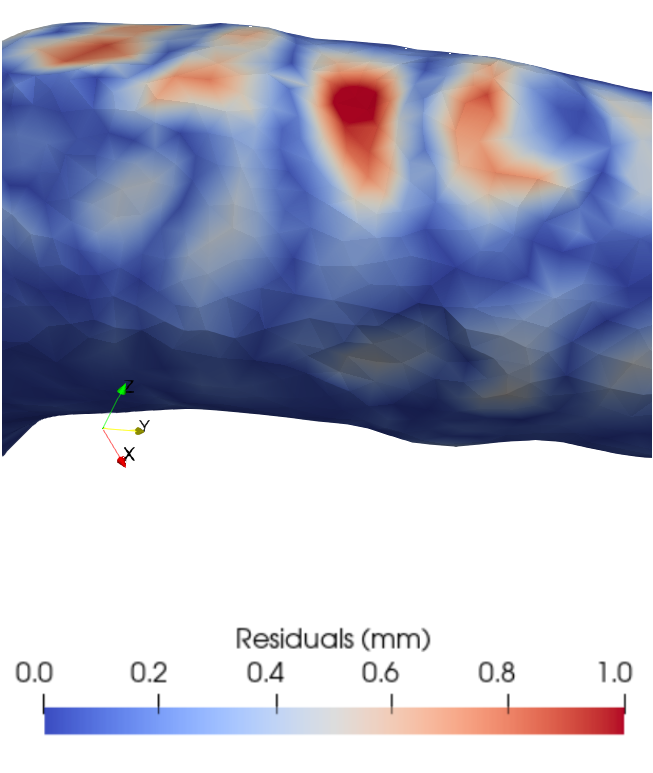}
\caption{On the first row, from left to right, the input aorta, the patient specific B-Spline approximation using $L=9,R=15,K=19$, and the superimposition of both (the approximation light blue). On the second row, the distribution of residuals (in mm) on the input mesh, with detailed views of the sinuses of Valsalva and the aortic arch, where the approximation is less accurate.}
\label{fig:residuals_mesh}
\end{figure*}

        The approximation of the wall is a smooth surface that can reproduce very closely the input data. In Fig.~\ref{fig:residuals_mesh} we can find, 
        in the upper row, from left to right, the input triangle mesh for one of the aortas, the B-Spline approximation using  $L=9$, $R=15$ and $K=19$, and both geometries superimposed (the approximation in blue).
        The lower row of the figure shows the distribution of the residuals for that case, with two close-up images corresponding to the regions with less approximation precision; around the sinuses of Valsalva and the upper wall of the aortic arch. Note that the value of the residuals distributes uniformly except for the highlighted regions and that, even there, the highest value of the residuals is around 1mm.

        To test the convergence of the patient specific model and the effect of $L$, $R$, and $K$ on the residuals, we have carried out a sensitivity analysis. 
        For each aorta, $\ba^k$, we have computed the mean residual, $\bar r^k$, and, in order to detect the existence of regions where the approximation could be less precise, the 75$^{th}$ quantile of the distribution, $r^k_{75}$. Finally, we have averaged these two indicators across the samples in the cohort. Next, we present the results for values of $L$, $K$ and $R$ in the range from 5 to 19.
        
\begin{figure}[ht!]
\centering
\includegraphics[width=.9\linewidth, trim={2cm 5cm 0cm 8cm}, clip]{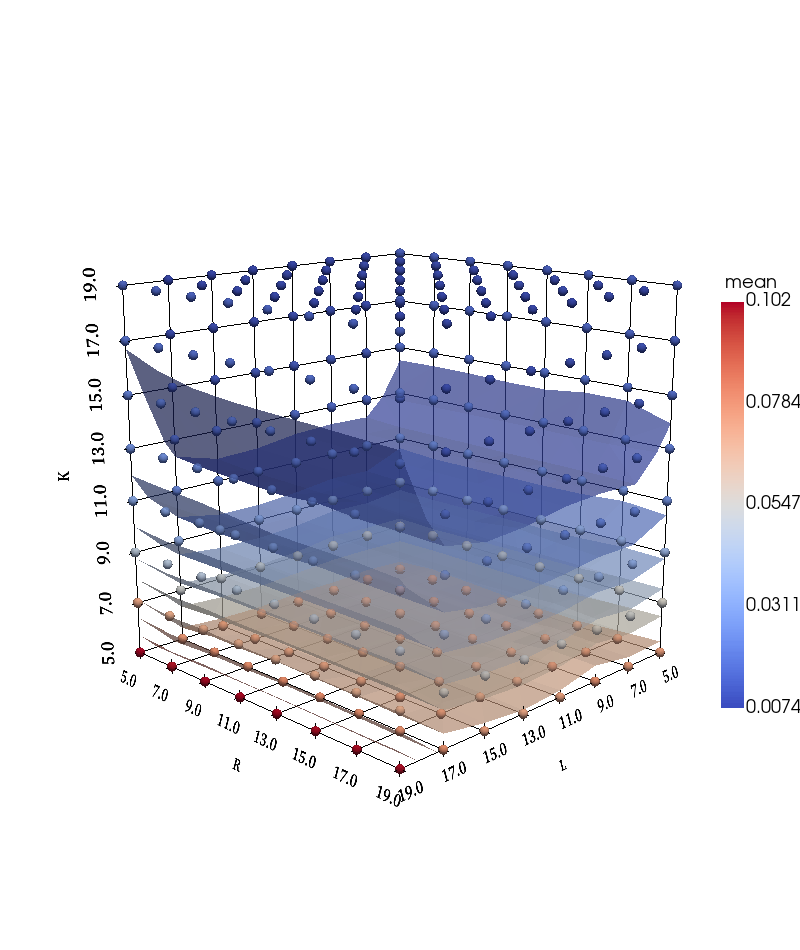}\\
\includegraphics[width=.9\linewidth, trim={2cm 5cm 0cm 8cm}, clip]{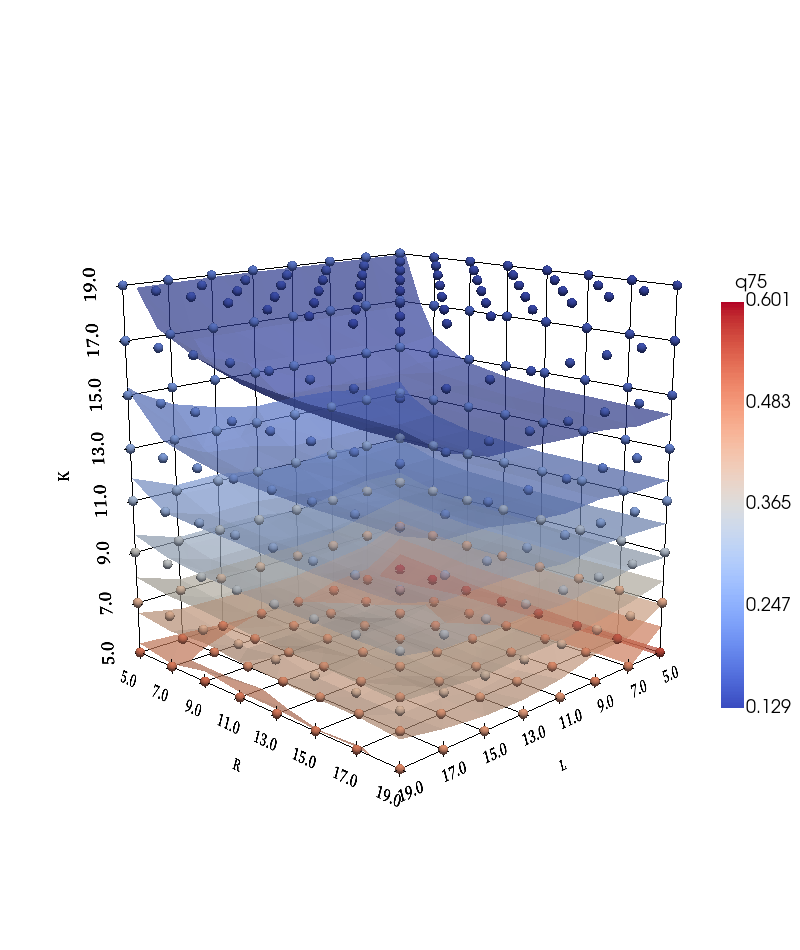}
\caption{Plot of the sensitivity analysis on the influence of the parameters $L$, $R$, and $K$ over the estimation residuals of the patient specific vascular model. The top plot shows the cohort average of the mean residuals, and the bottom plot shows the cohort average of the 75$^{th}$ quantile of the residuals, in both cases expressed in millimeters. In both plots, the dots placed in a regular grid are color-coded to indicate the value of the residual at the corresponding values of $L$, $R$ and $K$.}
\label{fig:residuals_sens_analysis}
\end{figure}

        In Fig.~\ref{fig:residuals_sens_analysis} the results of the analysis are presented in the form of isosurfaces of error, i.e., each surface represents the region in the space of parameters $(R,L,K)$ that correspond to a fixed value of the residual, encoded in the surface color. The results indicate that the parameter that has a highest influence in residual reduction is the number of knots for the surface in the longitudinal direction, $K$. Note that, the closer the surfaces are, the higher is the residual gradient with respect to the parameter in that direction. Thus, the surfaces that are closely stacked at the bottom of the plot imply that, when $K$ is low, an increase of knots in that direction yield higher error reduction. On the contrary, the rate of improvement decays as $K$ increases, indicating convergence.

        Given a value of $K$, the error is very stable w.r.t. the other two parameters, $R$ and $L$. This has a practical implication; given a value of $K$, which will govern the quality of the approximation, there is a reasonably wide range of values for both $L$ and $R$ that lead to a similar behavior. This means that, from the user perspective, tuning of the parameters is straightforward.
        
        We only observe an increase of error for very small or very large number of knots in the centerline approximation, $L$. The error at low values of $L$ is due to the lack of degrees of freedom to properly approximate $\cc$. On the contrary, when $L$ reaches $19$, an excessive adaption of $\cc$ to the sampling used to mark the centerline region (see Section~\ref{sssec:centerline_computation}) leads to a wiggling centerline curve, which translates into light oscillation on the wall. Nevertheless, when this effect is most noticeable, the error is still at its minimum values.

        The plot on the bottom of Fig.~\ref{fig:residuals_sens_analysis} corresponds to isosurfaces of error at the 75$^{th}$ quantile of the error distribution (averaged across the cohort). That is, it explains the behavior of the error in the regions corresponding to the upper tail of the error distribution. In this case, we see that low values of $R$ have a negative influence, specially at high values of $L$. Our interpretation is that, when the aforementioned wiggling effect appears in the centerline, small resolution in the traversal direction of the surface leads to a less accurate adaption to the observed data.

        \subsection{Population analysis}
        \label{ssec:SSM_results}

\begin{figure}[ht!]
        \centering
        \includegraphics[width=0.9\linewidth, trim={0cm 0cm 0cm 0cm}, clip]{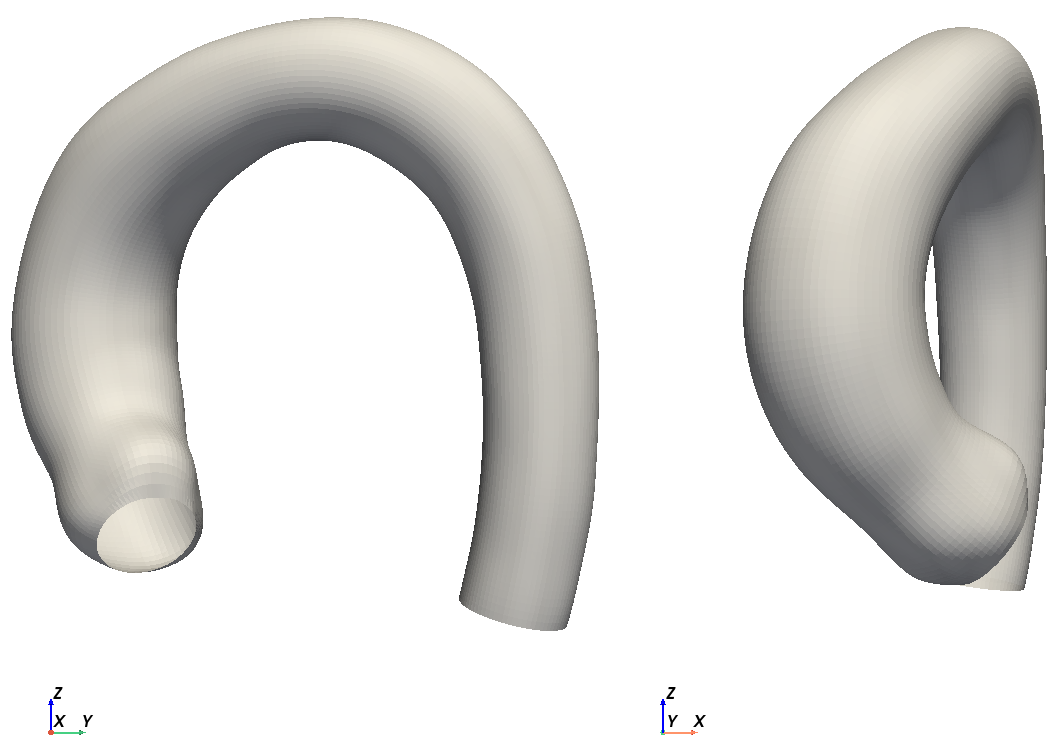}
        \caption{The saggital and coronal views of the mean thoracic aorta geometry obtained as a result of the PCA.
        }
        \label{fig:pca_mean_var}
        \end{figure}

        We have obtained the patient specific model for each of the 30 aortas in our cohort for $L=9$, $R=15$ and $K=19$. Then, we have performed a PCA on the resulting set of feature vectors, obtaining the mean aorta and the principal directions. Note that, if the number of points used during the PCA, $N$, is lower than the dimension of the space, the number of nonzero principal directions will be 
        {$N$, except for degenerate cases}. Thus, in our case, the PCA will yield 30 principal directions. Fig.~\ref{fig:pca_mean_var} shows the mean geometry obtained after the PCA.
        As it can be expected, the average geometry is a very smooth surface, without any particular protuberance. It is noteworthy that the sinuses of Valsalva have been captured in the mean aorta. This indicates that the VCS proposed is consistent across individuals, and that the derived point correspondence is properly defined.
                
\begin{figure*}[t!]
            \centering
            \includegraphics[width=\linewidth, trim={0cm 0cm 0cm 0cm}, clip]{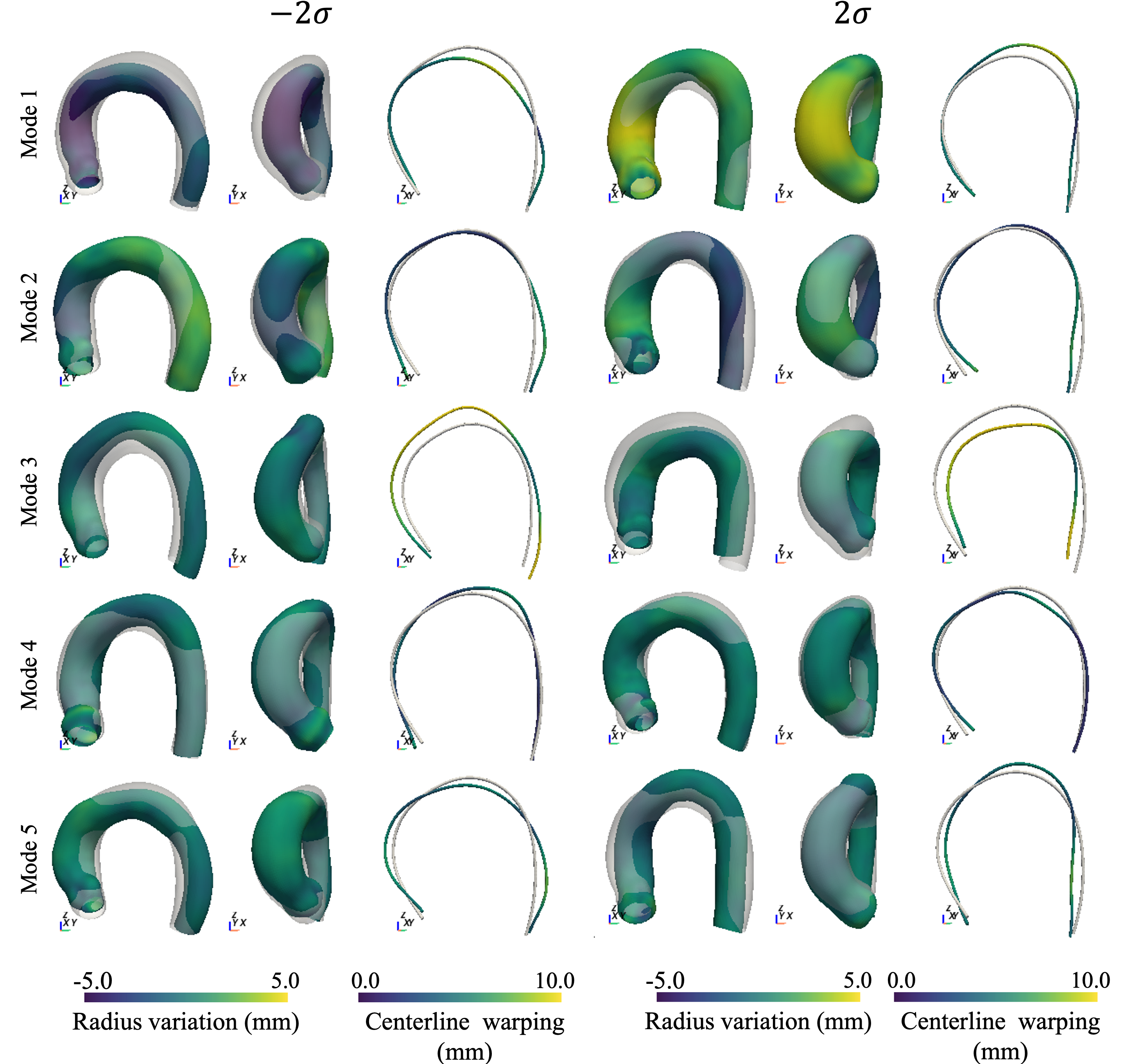}
            \caption{The illustration of the five deformation modes. Deformation modes, are sorted from top to bottom, then the first three columns are the negative deformation and the last three the postive . The deformation of each mode is split in the contribution of the centerline and the radius function. The gray phantom shape corresponds to the mean aorta in both surface and centerline.}
            \label{fig:pca_5comps}
        \end{figure*}

        The VCS allows a separate analysis for the centerline and the wall of the vessel.
        Fig.~\ref{fig:pca_5comps} presents a graphical comparison of the first 5 deformation modes obtained by the PCA. For each deformation mode, $\uu_i$, the figure shows the wall of the aortas $\bmu \pm 2\sigma_i\uu_i$, where $\sigma_i$ is the observed variance of the cohort along that direction. Each deformed wall is shown in saggital and coronal view, superimposed with a phantom of the mean aorta, for comparison. 
        In addition, the deformed centerline is also shown compared to the centerline of the mean aorta.
        Using the one to one correspondence between points on two vessels, defined by the VCS, differences between the mean shape an the deformed anatomy have been computed and represented using a color scale. 
                
        In the case of the wall, the difference in the radius of the aorta is represented for each point. This shows, e.g., how the first deformation mode is responsible for a narrowing or widening of the whole lumen, while the second mode of variation has an opposite effect on ascending and descending sections. The other three modes represented have more local effects, as in the fourth mode where the main variations can be observed in the sinuses of Valsalva and the sinotubular junction, or in the fifth mode, that affects the ascending aneurysm anatomy.
        
        The warping of the centerline has been measured as the $\bbbr^3$ euclidean distance between points with the same value of $\tau$ on the mean and the deformed curves. The representation of the deformation enables an easy inspection of the regions affected by each deformation mode. This way, we see how the third mode of deformation has a rather uniform effect on the whole curve, while the rest of modes cause more local displacements. It has to be noted that the distances in the plot can appear smaller than they actually are due to the projection.
        This analysis shows that, in our cohort, the variation of shape can be described mainly by the deformation of the distal part of the aortic arch (first deformation mode) and by the flatness of the arch (second mode of variation) and by the vertical roundness of the global shape (fifth mode of variation). It is also noteworthy that two of the five deformation modes (the second and fourth modes) have little impact on the centerline shape, with effect mainly on the wall radius.

        \subsection{Haemodynamic atlas}
        \label{ssec:haemo_atlas_res}
                
\begin{figure}[ht!]
        \includegraphics[width=.08\linewidth, trim={0cm 0cm 0cm 0cm}, clip]{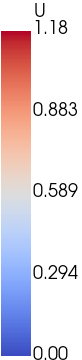}\hfill
        \includegraphics[width=.9\linewidth, trim={0cm 3.0cm 0cm 5.6cm}, clip]{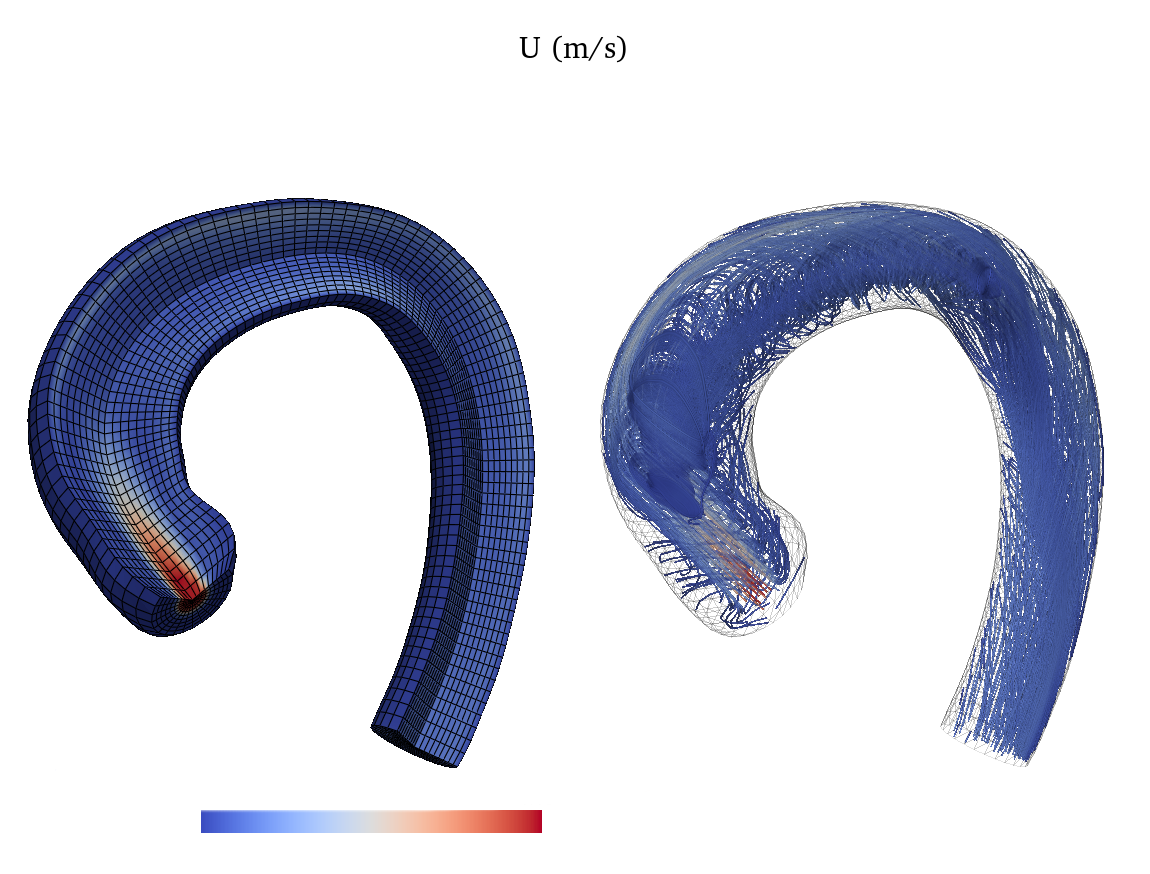} \\
        \includegraphics[width=.08\linewidth, trim={0cm 0cm 0cm 0cm}, clip]{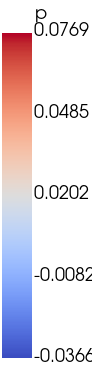} \hfill{}
        \includegraphics[width=.9\linewidth, trim={0cm 3.0cm 0cm 5.6cm}, clip]{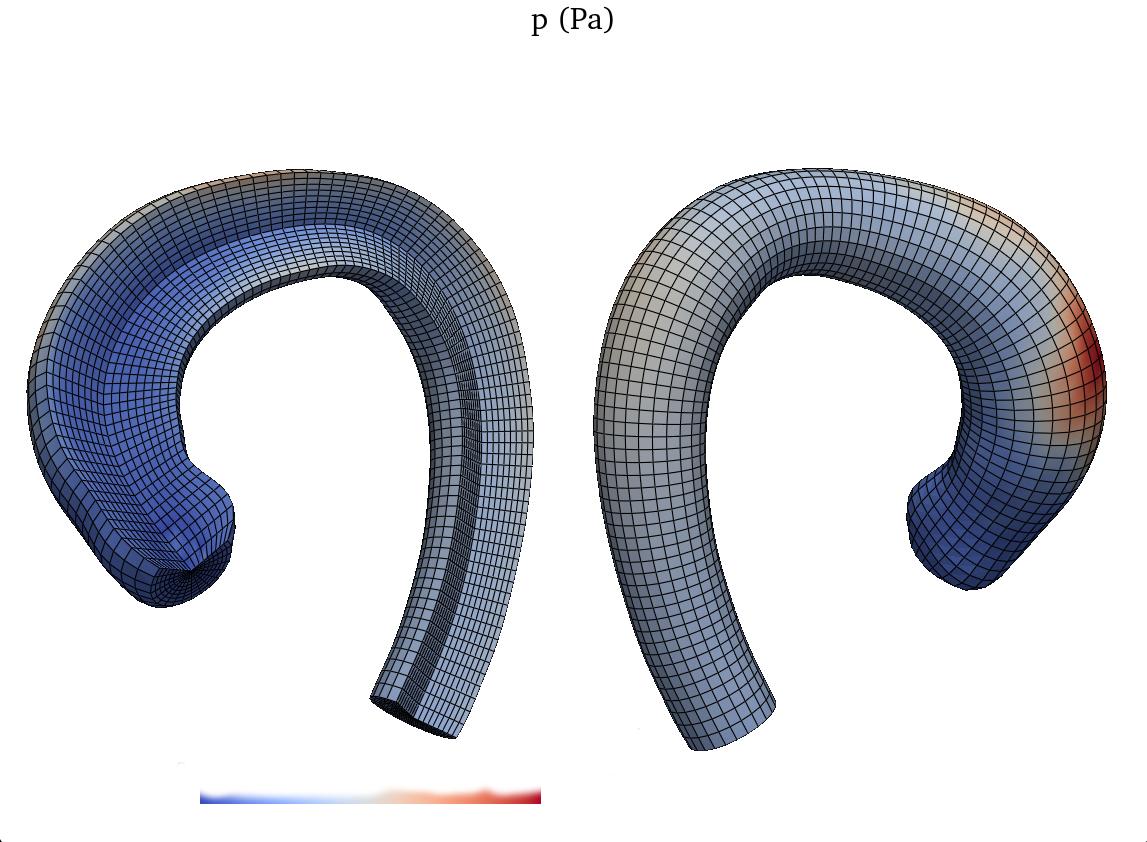} \\
        \includegraphics[width=.08\linewidth, trim={0cm 0cm 0cm 0cm}, clip]{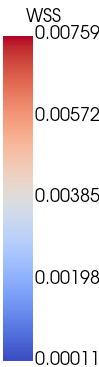} \hfill{}
        \includegraphics[width=.9\linewidth, trim={0cm 3.0cm 0cm 5.6cm}, clip]{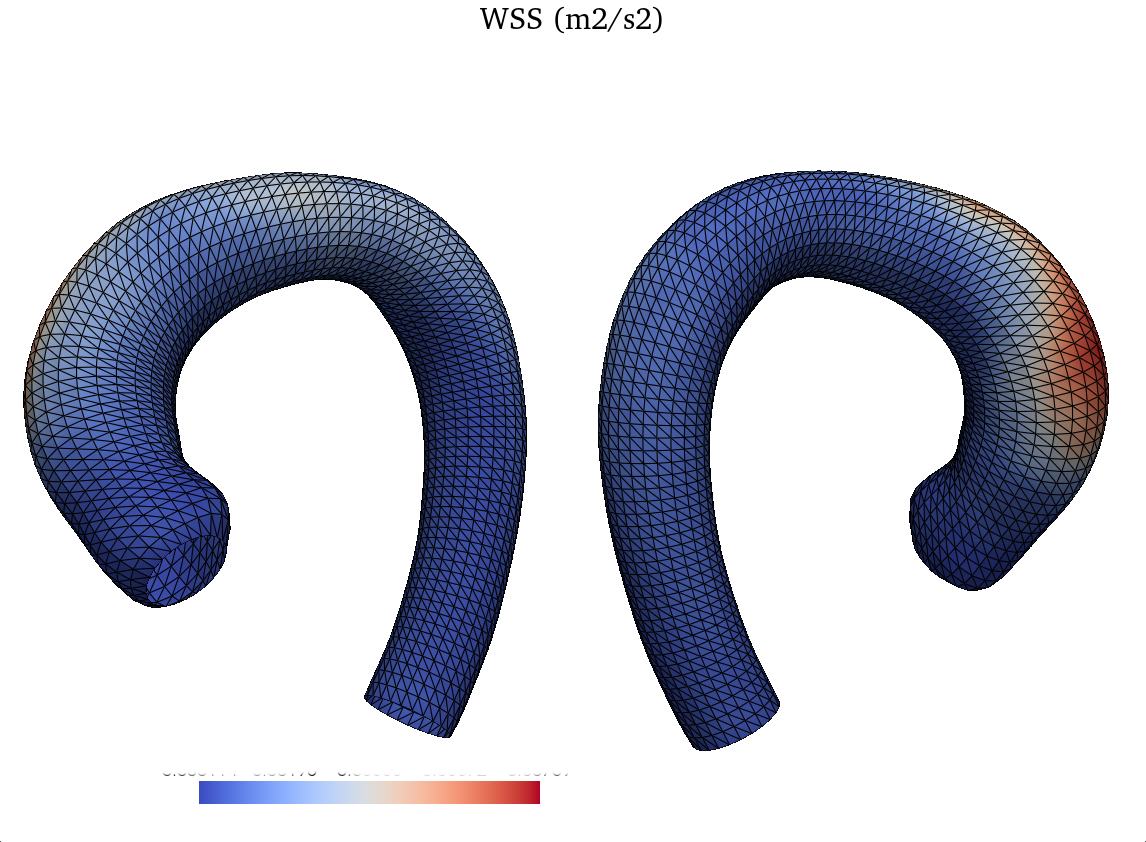} 
        \caption{Using the VCS, a grid is used to evaluate the flow on all the aortas of the cohort and, then averaged. 
        The result is represented over the average anatomy. From top to bottom, velocity (m/s), pressure (Pa) and wall shear stress (m$^2$/s$^2$). 
        The internal mesh corresponds to the measurement points, and not to the computational grid used for CFD simulation.}
        \label{fig:haemo_pca}
        \end{figure}

        A CFD simulation has been performed on every aorta using the finite volume method, as described in Section~\ref{sssec:haemo_atlas}.
        Fig.~\ref{fig:haemo_pca} shows the average fields, measured on the cohort, for the three quantities inspected.
        {Velocity is plotted, in the first row of the figure, over a regular internal grid on the mean geometry obtained in Section~\ref{ssec:SSM_results}. The left figure presents the magnitude of the vector field. To show the interior of the fluid, the volume is clipped using a threshold in the coordinate $\theta$ of the aorta lumen, and the value is shown on the resulting surfaces. This kind of plot, which cannot be done by merely clipping the volume using planes, becomes extremely easy thanks to the VCS.
        In the right figure, the first row presents streamlines integrated along this field. In a similar fashion, the second row presents the value of pressure over the same grid. In this case, the right plot presents the value of pressure on the wall from a posterior view. Finally, the third row presents wall shear stress, which is only defined on the wall, from the anterior and posterior views.}


        As in the case of the anatomy, where the sinuses of Valsalva were captured by the mean geometry, the mean flow presents traits that are characteristic to the flow of the individual aortas. This, again, points towards a proper point correspondence between individuals that helps to capture the characteristics of the flow at each different region of the aorta.

\begin{figure*}[ht!]
            \SetTblrInner{rowsep=0pt}
            \begin{tblr}{Q[l,h]Q[c, b]Q[r, b]}
            \includegraphics[scale=0.25]{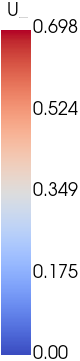} &
            \includegraphics[width=.44\textwidth, trim={.5cm 5.5cm 0cm 5cm}, clip]{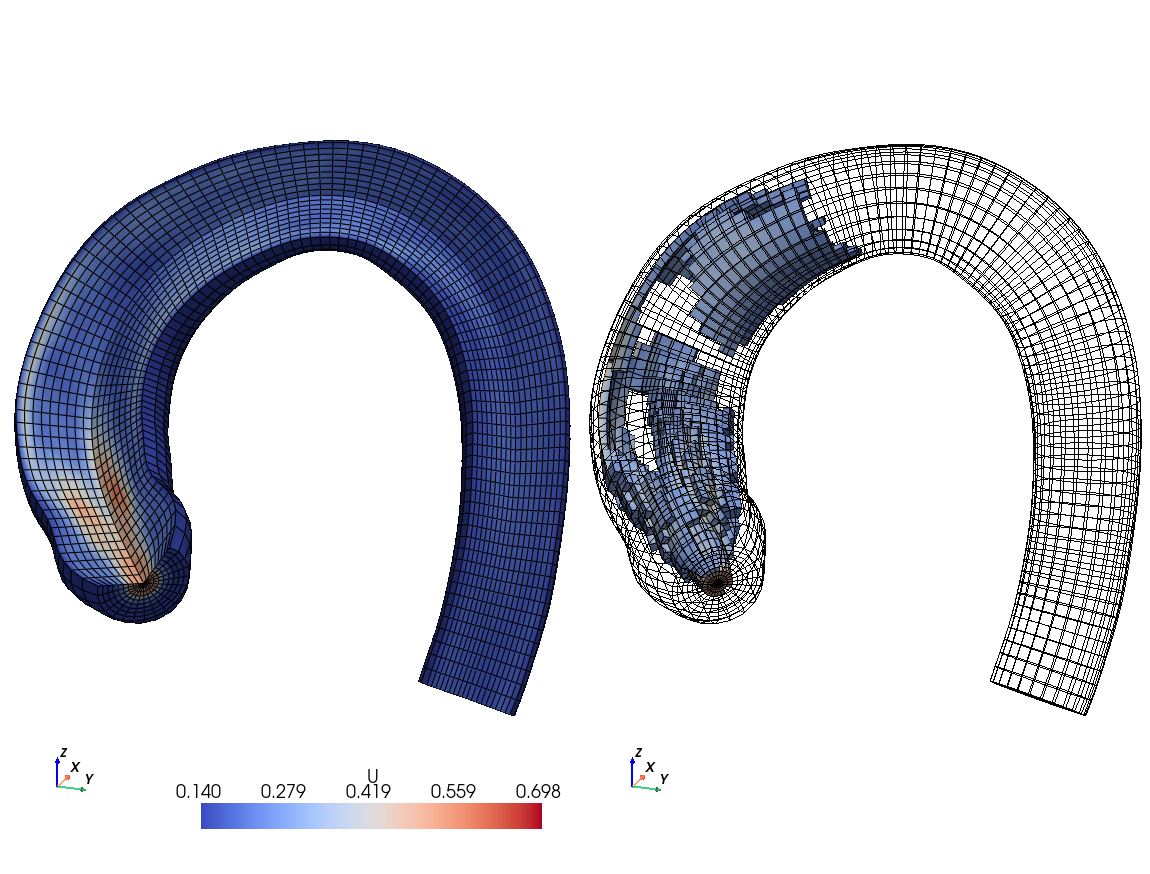} &
            \includegraphics[width=.44\textwidth, trim={.5cm 5.5cm 0cm 5cm}, clip]{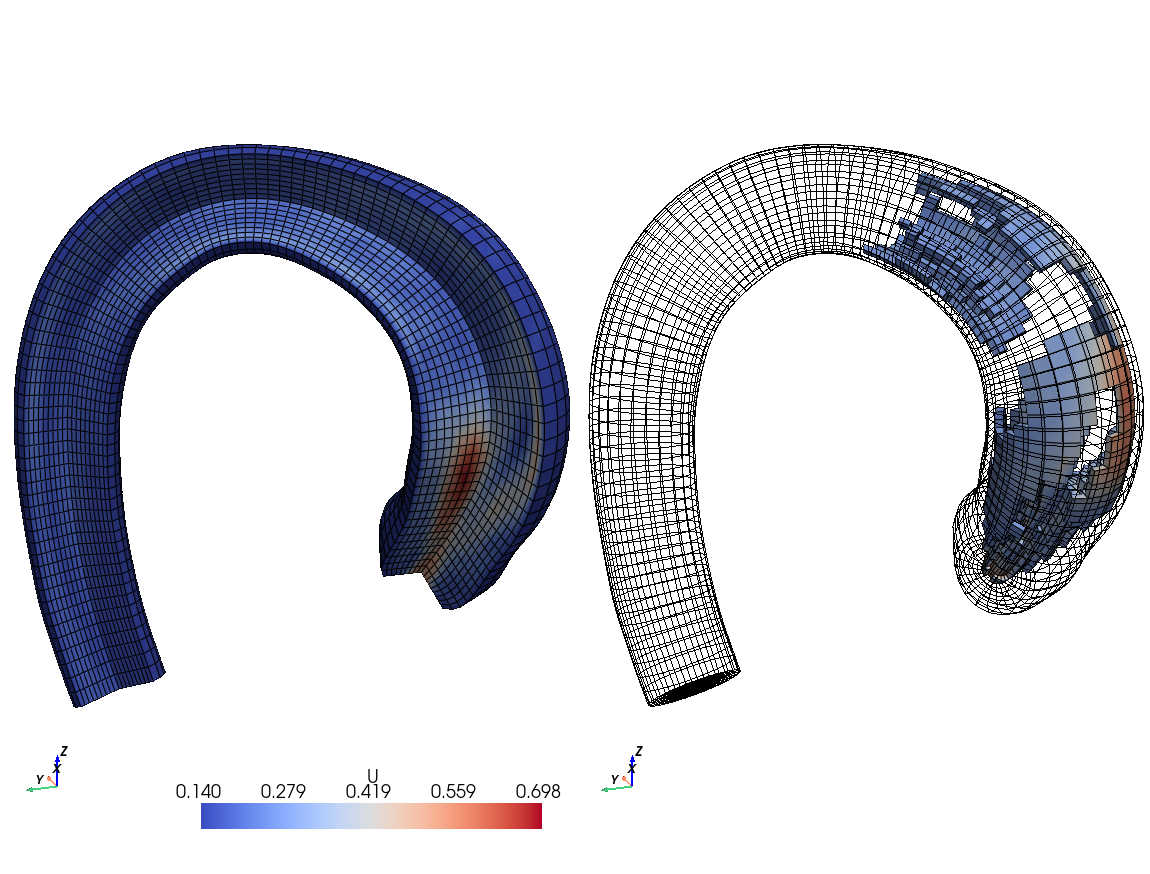}\\
            \includegraphics[scale=0.25]{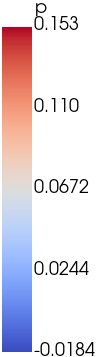} &
            \includegraphics[width=.44\textwidth, trim={.5cm 5.5cm 0cm 5cm}, clip]{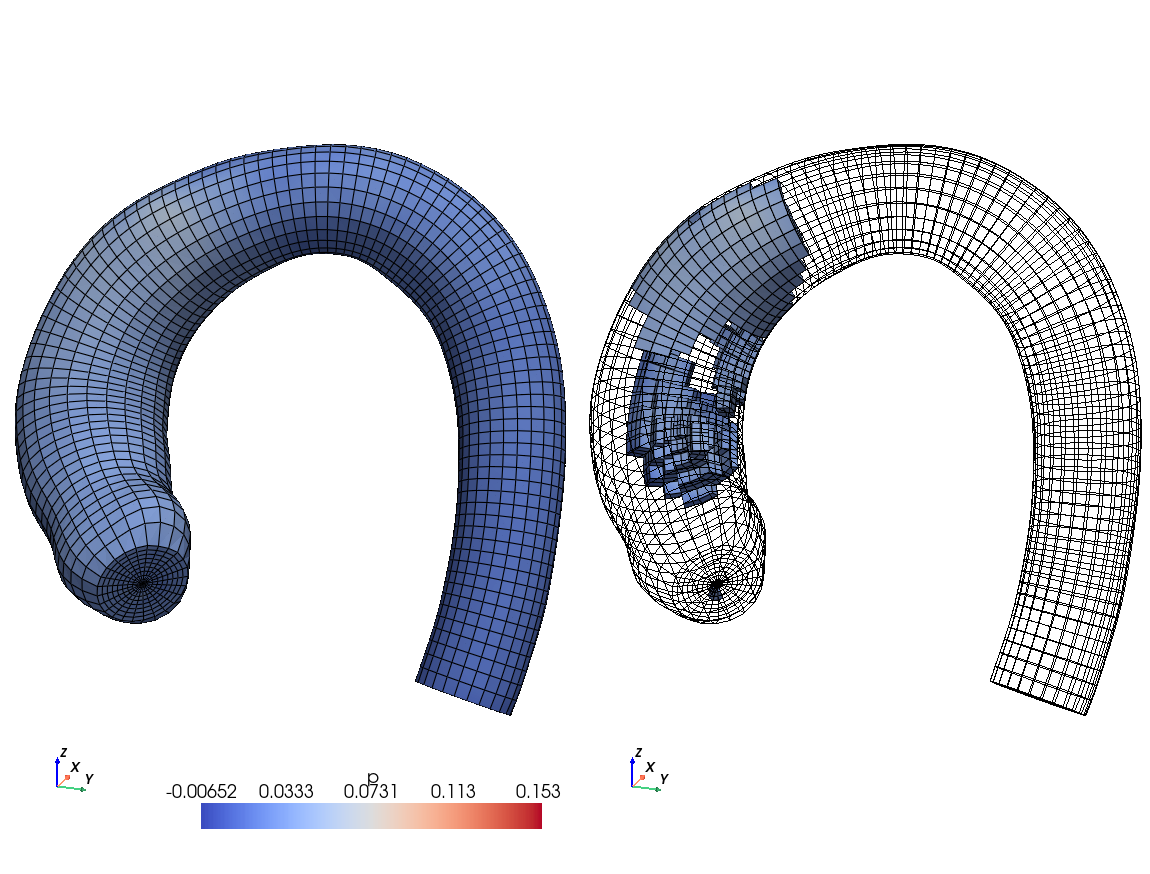} &
            \includegraphics[width=.44\textwidth, trim={.5cm 5.5cm 0cm 5cm}, clip]{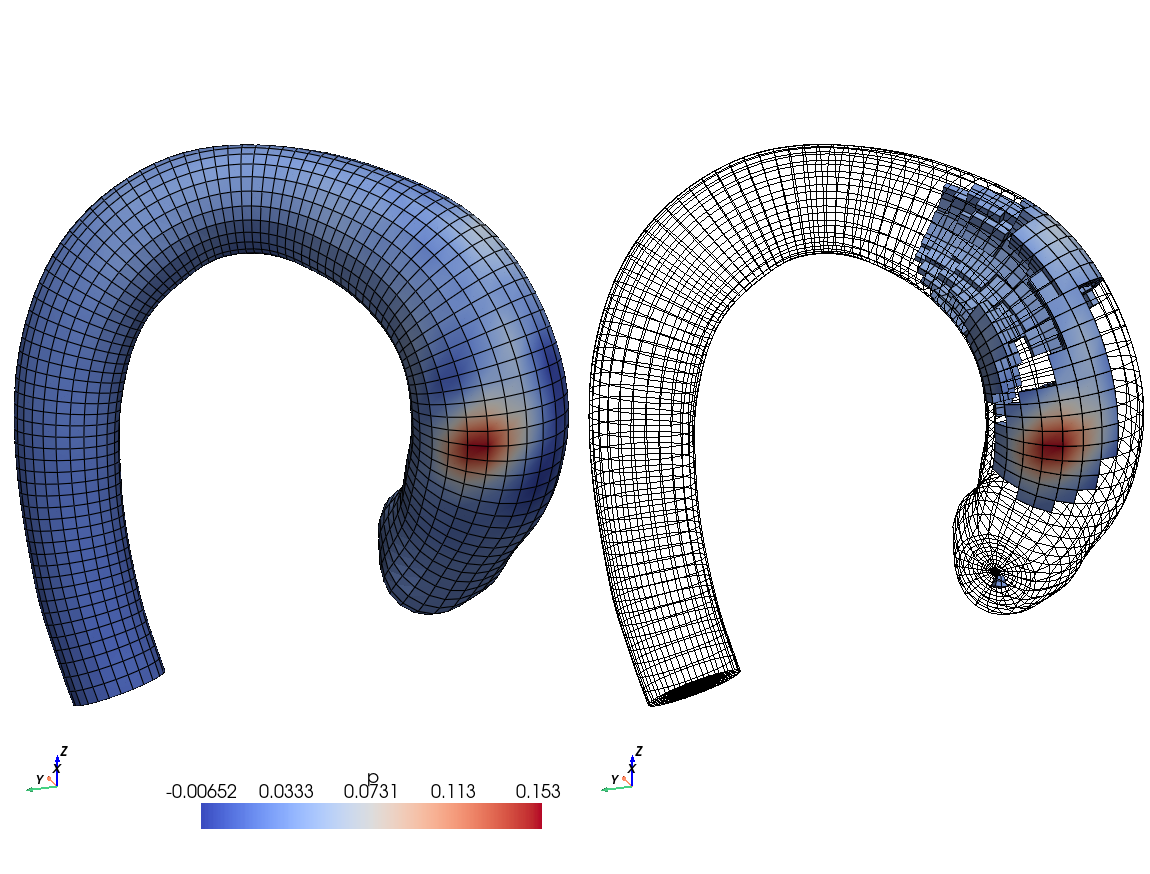}\\
            \includegraphics[scale=0.25]{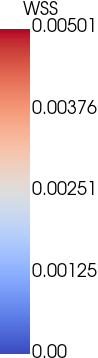} &
            \includegraphics[width=.44\textwidth, trim={.5cm 5.5cm 0cm 5cm}, clip]{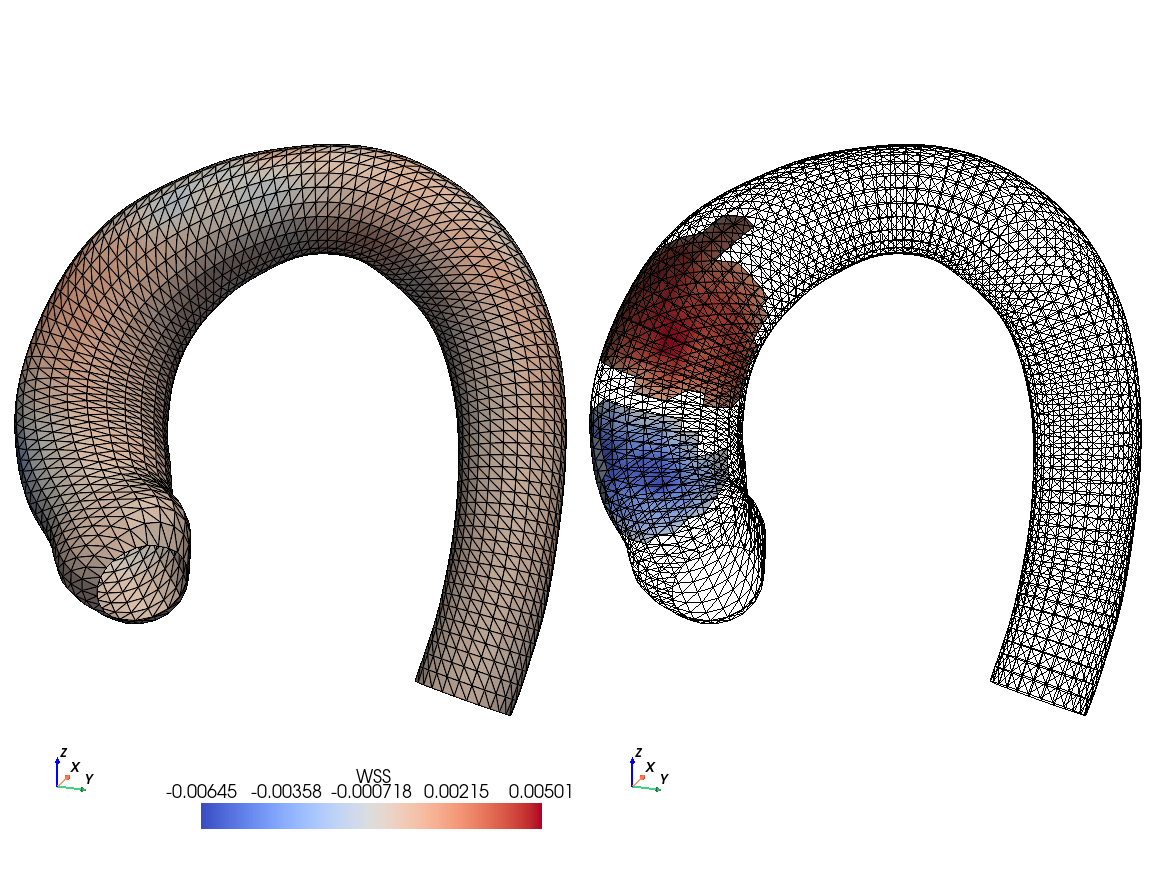}&
            \includegraphics[width=.44\textwidth, trim={.5cm 5.5cm 0cm 5cm}, clip]{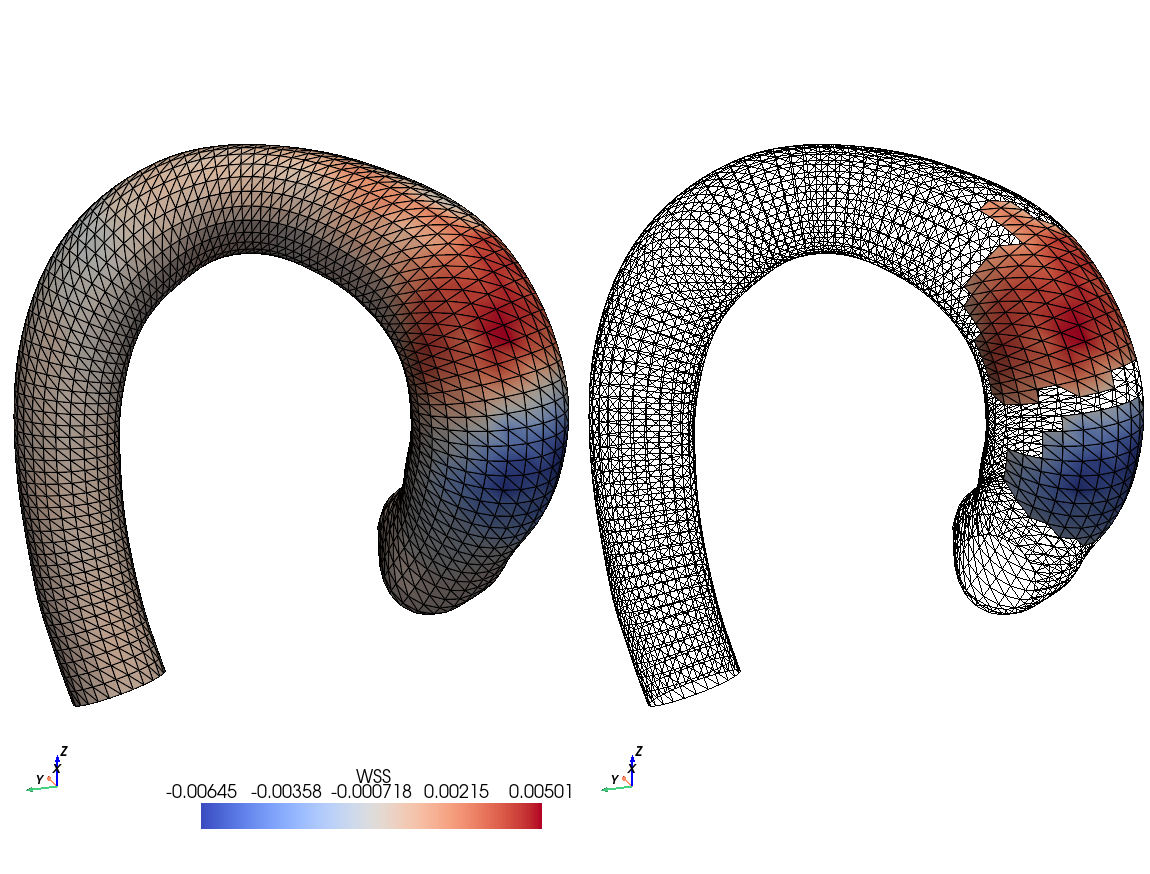}
            \end{tblr}
            \caption{From top to bottom, the first principal component of the velocity, U, (m/s), pressure (Pa) and wall shear stress (m$^2$/s$^2$). The solid mesh for velocity is  displayed open to show the interior of the lumen. The second and fourth columns present the regions where the variation is at least a 20\% from the total maximum variation. The meshes used in the figures are not the CFD computational meshes but a regular grid built using the VCS for visualization.}
            \label{fig:haemo_pc}
            \end{figure*}

        The use of the VCS as a tool for measuring and visualizing data in a vessel also helps to inspect the variation of the flow within the cohort. In Fig.~\ref{fig:haemo_pc}, the first mode of variation of the three flow fields of interest is shown. In this case, the plot shows
        the first principal component multiplied by $2\sigma_1$. Only the positive increment has been included, since the flow difference using $-2\sigma_1$ would simply invert the colors of the plot. 

        The figure is presented in three rows and four columns; each row corresponding to a variable of the flow; and the two first columns correspond to ventral view, while the two later columns correspond to dorsal view of the aorta.
        In first and third columns, the value of flow is represented in the regular mesh built using the VCS for field evaluation. In the case of the velocity (first row) the grid is clipped using the coordinate $\theta$, to enable inspection of the lumen, as in Fig.~\ref{fig:haemo_pca}.
        In the second and fourth columns, the mesh is clipped using the value of the variable of interest. That is, the cells with a value in the upper 20\% of the observed range are shown, to highlight which are the main areas of influence of the the principal component. These regions correspond, thus, to the regions of the aorta in which higher variability is shown across the cohort.


\section{Discussion}
\label{sec:discussion}
We have defined a coordinate system that adapts to vessel anatomy, and allows the unambiguous identification of a point in both the lumen and the wall. This enables the definition of a correspondence between the anatomy of different patients, with applications to inter-patient comparison, quantitative cohort analysis or patient specific data aggregation.
The computation of the VCS and the derived methods only requires the identification of the vessel segment of interest and the definition of a reference frame for one of the two ends. Thus, once the region of interest has been segmented, it is a completely automatic process which takes less than a couple of minutes for a thoracic aorta on a standard desktop PC.

The coordinates of a point are well defined as long as Eq.~(\ref{eq:vcs_tau}) has a unique solution, which is met for the set $\Omega$ of the points that are closer to $\cc$ than the curvature radius of the centerline. Thus, the VCS can be computed in the region of points that are close enough to the curve according to this criterion.
In the vast majority of cases, this region will include the complete vessel lumen. 
The VCS has been tested on the aortic arches of a cohort of patients with ascending aorta aneurysm. This scenario includes large vessel radius sections, and, in all the cases, the whole vessel was within $\Omega$. Nevertheless, when computing the centerline this condition can be forced during the application of the A$^*$ algorithm (Section~\ref{sssec:centerline_computation}).

The proposed coordinate system generalizes other comparable approaches that use cylindrical coordinates for vessels. \cite{bersi16, bersi19} use a straight line segment as the longitudinal direction, while \cite{meister20} rely on the Frenet frame for the centerline. 
This limits their scope to work with nearly straight vessel segments, which is an impractical requirement at large regions of the vascular system.
The generalization of the longitudinal axis by a longitudinal curve, and the parallel transport frame used in our proposal allow to overcome this limitation. 
Our approach also generalizes the concept of canal surfaces, that consider circular or elliptical cross sections~\citep{alvarez17,antiga08,piccinelli09}, allowing the construction of accurate patient specific models. \cite{guo13} and \cite{ghaffari17} propose comparable approaches, but limit their work to mesh generation, while our definition of vessel coordinates enables inter- and intra-patient point to point correspondence.

The patient specific vascular model built upon the VCS involves a very reduced number of parameters ($L$, $K$ and $R$) which, according to our tests, are easy to choose; 
we have found that the approximation error for our cohort is clearly below the usual medical imaging resolution, and that the quality of the adjustment can be controlled just by increasing $L$ for a good range of values of the other parameters.
Our approach is less general than other strategies that serve for arbitrary surfaces~\citep{bruse16,bruse17,sophocleous18}, but its computation is very fast, taking barely a few seconds once the VCS has been obtained. In addition, the representation based on B-Splines coefficients is very intuitive thanks to its geometrical interpretation.

The model is low dimensional, compared to the equivalent medical image or triangle mesh surface data; a few hundred floating point values are enough to achieve a very good approximation for the aorta anatomy. All the samples in a cohort have the same dimension, with comparable feature vectors, which is most often a requirement for any statistical or machine learning analysis~\citep{bruse17,liang20,meister20,niederer20,sophocleous18}.
Using this property, the anatomical variability of our cohort is analyzed by means of a PCA on the set of patients' feature vectors of B-Spline coefficients, as shown in Section~\ref{ssec:SSM_results}.
{
Note that this analysis can also be done for different CT or MRI scans of a single patient. This would enable temporal analysis of cardiac cycle or assessment of disease evolution along time, e.g., to evaluate aneurysm growth.
}

Our results are consistent with other shape variability analyses, indicating the importance of modes that affect the diameter of the aorta and the arch shape~\citep{casciaro14,catalano21,ou07}. In addition, we use the point-to-point correspondence on the surface of the vessels, and the explicit consideration of the centerline in the patient specific model, to perform a more insightful analysis.
When a deformation is represented combining the arch and wall displacements, it can be difficult to clearly understand how much of the displacement comes from the arch shape distortion and how much is due to diameter changes. However, we are able to show the arc shape variation with the deformed centerline and quantify the amount of wall displacement w.r.t. that centerline in the resulting shape, as shown in Fig.~\ref{fig:pca_5comps}. This helps a better identification of the regions that are actually affected by diameter changes.
{
This ability to separate the variation of shape in the centerline and in the wall diameter can give insight on how anatomy at different scales correlates with clinical condition~\citep{wu19,lovato24}.
}

Geometric consistency across patients has been reported as a relevant property for anatomy representation, both for obtaining valid large deformations~\citep{catalano21} and for avoiding spurious deformations in the principal variation modes~\citep{thamsen21}.
It is remarkable that, in the mean aorta, the sinuses of Valsalva are observable, indicating the inter-patient consistency of the vessel coordinate system.
The mean flow of the haemodynamic atlas shown in Section~\ref{ssec:haemo_atlas_res} points in the same direction, since features such as the helicity along the aortic arch or the WSS profile in the ascending aorta aneurysm have not been smoothed away due to coordinate mismatches between patients.

The haemodynamic atlas built in Section~\ref{ssec:haemo_atlas_res} differs from that of~\cite{catalano21} in the sense that they run their CFD simulations on the shapes resulting from the deformation modes, while we perform them on the original cohort. \cite{thamsen21} also run the simulations on the cohort, but they do not perform a variability analysis. Instead, they compare pressure measured along the centerline. We exploit the VCS to build an internal measurement mesh, equivalent for all the aortas, and apply a PCA on the resulting numerical values.
This provides powerful visualization approaches, as shown in Figs.~\ref{fig:haemo_pca} and~\ref{fig:haemo_pc}. 

{
            In our CFD setup, the wall was considered rigid. 
            However, if fluid-structure interaction (FSI) is used to account for wall deformation, the measurement mesh will automatically adapt to the deformed shape, requiring no additional computation.
            Moreover, in FSI simulations, wall representation as a polynomial surface enables the application of isogeometric analysis~\citep{zhang07} and can help describe heterogeneous wall properties by describing them in the $\tau-\theta$ space.
}

Moreover, the proposed representation for anatomy and flow is well suited for its use in statistics and machine learning methods~\citep{liang20}, enabling quantitative analysis in the complete lumen and the wall.
In a very preliminary analysis, we have found the ability to predict some flow variation modes from the first 20 deformation modes using linear regression. As expected, pointwise prediction of the flow has not been achieved for the whole aorta, but, remarkably, there exists regions in which even linear regression still yields good results.

The presented methodology to encode flow fields offers a new approach to diverse problems, such as the combination of data from different sources, for instance combining 4D flow MRI with numerical simulations, 
{to impose boundary conditions}, to validate in both directions, or to tackle the denoising problem~\citep{fathi20}.
{
VCS coordinates on the wall can also be used to describe the distribution of mechanical properties if tissue or wall thickness, taking the same approach applied to build the haemodynamics atlas.
}
Moreover, the standard formatting of the encoding makes it appropriate as the input for machine learning techniques. With a different encoding method this approach was used by~\cite{liang20}, where the authors predicted the main flow fields and analyzed the feasibility of using said predictions in clinical application.

In the context of the development of the digital twin, our proposal can serve as a robust and versatile vessel encoding method, that covers both anatomy and biophysical data. Although not tested in this work, its utility for aggregating data from different sources is clear, and we also think that it can help to find anatomical landmarks. Moreover, the fact that the description proposed is differentiable makes it adequate for differentiable programming~\citep{xie23} or uncertainty quantification~\citep{perinajova21,rego21}.

A limitation of the VCS is that it cannot deal with vessel branching.
%
{
Vessel bifurcations are regions of great interest, since atherosclerotic plaques develop around them and they can have great impact on flow dynamics.
}
Nevertheless, this problem can be overcome, with some effort, by applying the VCS to each vessel section and building a vessel tree with existing methods for branching description~\citep{antiga04,ghaffari17,guo13,medrano-gracia16}.
The election of uniform B-Splines helps to establish the point correspondence between vessels. This, however, can lead to artifacts if the vessels have several sections with specific characteristics but with variable length. A proper reparametrization of $\cc$ can help in this situations, by imposing certain values of $\tau$ at the landmarks of interest.

\section{Conclusions}
\label{sec:conc}

We have defined a vessel coordinate system that allows the unambiguous identification of locations in a vessel segment. We have described the technical details necessary for its computation, discussing the main assumptions taken and how they guarantee a proper definition. To show the applications of our proposal, we have analyzed the anatomical variability of a cohort of thoracic aortas and built an haemodynamic atlas. The VCS has been used at different stages to enable geometry encoding, quantitative comparison, flow variability analysis and advanced visualization. Although some further advancements are yet to be developed, specially to introduce branching structures, the proposed methods could serve as a robust piece for vascular modeling in the development of the digital twin.



\section*{Author contributions}
PR, ML and IGF Conceptualized the vessel coordinate system, its applications and the code involving its computation. PR and AP, designed and carried out the blood flow simulations. PR, ML and IGF designed and developed the code involving the visualization and graphic material. PR, RS, ML and IGF  wrote the original draft. RS and IGF acquired the funds to carry out the work.

\section*{Conflict of interest statement}

The authors declare that they have no known competing financial interests or personal relationships that could have appeared to influence the work reported in this paper.

\section*{Acknowledgements}

This work was funded by Generalitat Valenciana Grant AICO/2021/318 (Consolidables 2021) and 
Grant PID2020-114291RB-I00 funded by MCIN/ 10.13039/501100011033. 
None of the sponsors were involved in any step of the execution of the study or the preparation of the manuscript.
The authors would like to thank Laia Romero for her advice in the generation of graphical material.


\bibliographystyle{plainnat} 
\bibliography{vcs_doi}

\end{document}